%% file: paper_ieee.tex
\documentclass[10pt,journal]{IEEEtran}
\usepackage{amsmath}
\usepackage{amsfonts}
\usepackage{algorithm}
\usepackage{graphicx}
\usepackage{algpseudocode}
\usepackage[utf8]{inputenc}
\usepackage[T1]{fontenc}
\usepackage{url}
\usepackage[binary-units=true,per-mode=symbol]{siunitx}
\usepackage{paralist}
\usepackage{footnote}
\usepackage{caption}
\usepackage[bottom]{footmisc}
\usepackage{array}
\usepackage{subfigure}
\usepackage{nicefrac}
\usepackage{acronym}
\usepackage{setspace}
\usepackage[shortlabels]{enumitem}

\AtBeginDocument{%
	\addtolength\abovedisplayskip{-0.17\baselineskip}%
	\addtolength\belowdisplayskip{-0.17\baselineskip}%
	\addtolength\abovedisplayshortskip{-0.17\baselineskip}%
	\addtolength\belowdisplayshortskip{-0.17\baselineskip}%
}

	\newacro{ND}[ND]{neighbor discovery}
	\newacro{MANET}[MANET]{mobile ad-hoc network}
	\newacro{BLE}[BLE]{Bluetooth Low Energy}

\begin{document}
	
\title{Optimizing BLE-Like Neighbor Discovery}
\author{
Philipp H. Kindt, Swaminathan Narayanaswamy, Marco Saur, \textit{Technical University of Munich (TUM), Germany}\\
Samarjit Chakraborty, \textit{University of North Carolina at Chapel Hill, USA} \vspace*{-0.5cm}}
\noindent To appear in the IEEE Transactions on Mobile Computing.
Author’s version September 2020.
The published version might be different. \\
\\
© 2020 IEEE. Personal use of this material is permitted. Permission from IEEE must be obtained for all other uses, in any current or future media, including reprinting/republishing this material for advertising or promotional purposes,creating new collective works, for resale or redistribution to servers or lists, or reuse of any copyrighted component of this work in other works.
\maketitle
	\begin{abstract}
		\input{sections/abstract.tex}

	\end{abstract}

\input{sections/content.tex}

%\newpage
\vspace*{-0.5em}
\bibliographystyle{IEEEtran}
\bibliography{literature}
\vspace*{-1.0cm}

\begin{IEEEbiography}[{\includegraphics[width=1in,height=1.25in,clip,keepaspectratio]{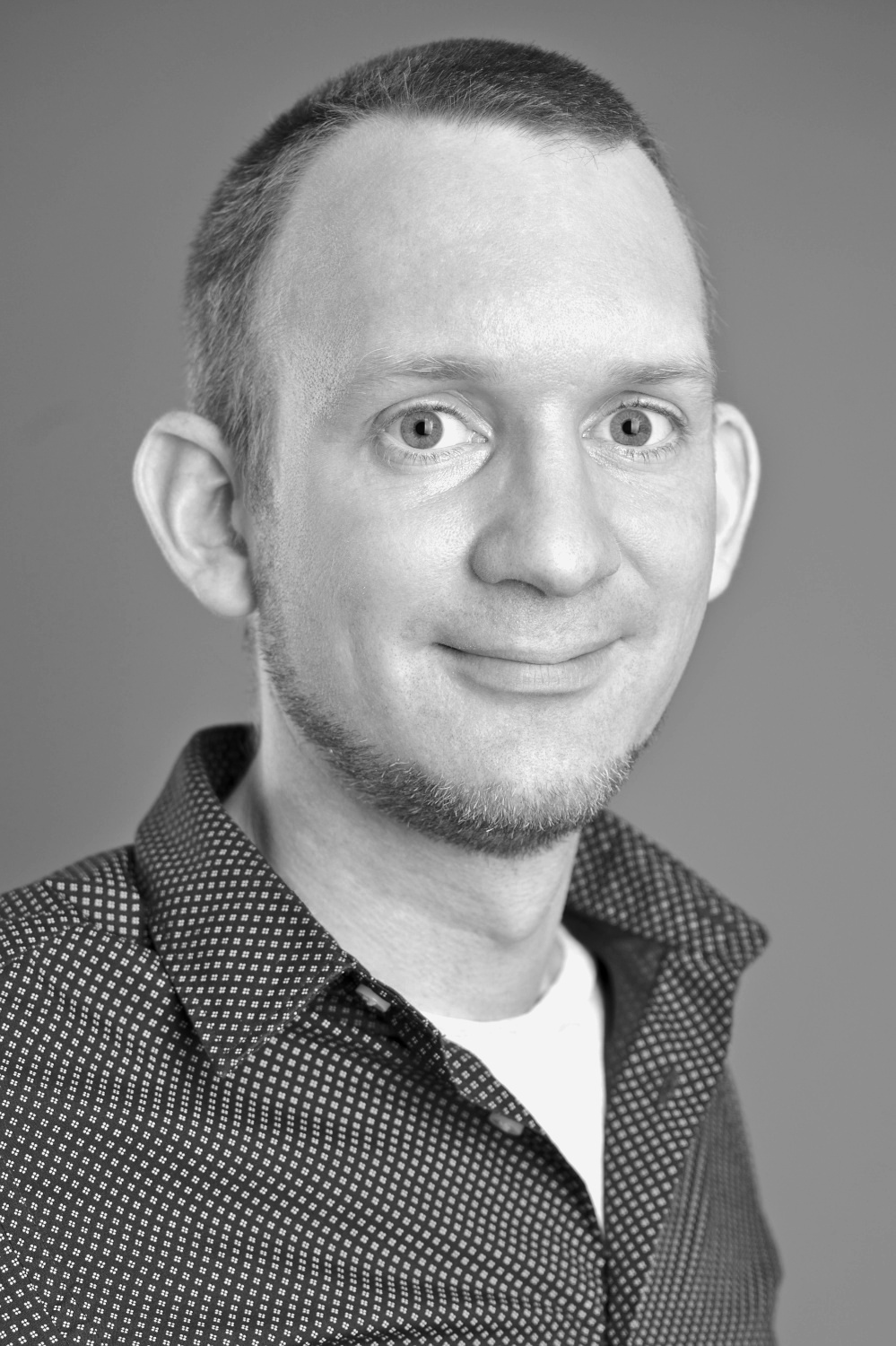}}]{Philipp H. Kindt} received his Ph.D. in Electrical Engineering from Technical University of Munich (TUM) in 2019. He has authored multiple peer-reviewed papers at conferences such as ACM SIGCOMM, IEEE INFOCOM or ACM/IEEE IPSN. His research interests are wireless communication, mobile computing and the IoT.
\end{IEEEbiography}
\vspace*{-1.1cm}
\begin{IEEEbiography}[{\includegraphics[width=1in,height=1.25in,clip,keepaspectratio]{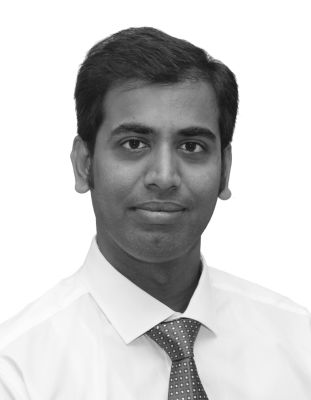}}]{Swaminathan Narayanaswamy}
	Swaminathan Narayanaswamy received his M.Sc. degree in IC design from Technical University of Munich (TUM) and Nanyang Technological University, Singapore. He obtained his Ph.D. degree from TUM, where he is currently working as a senior research fellow. 
	Before joining TUM in 2017, he worked as a research fellow at TUM CREATE in Singapore.
	His main research interests are DC-DC converters, battery management systems and cell balancing.
\end{IEEEbiography}
\vspace*{-1.1cm}
\begin{IEEEbiography}[{\includegraphics[width=1in,height=1.25in,clip,keepaspectratio]{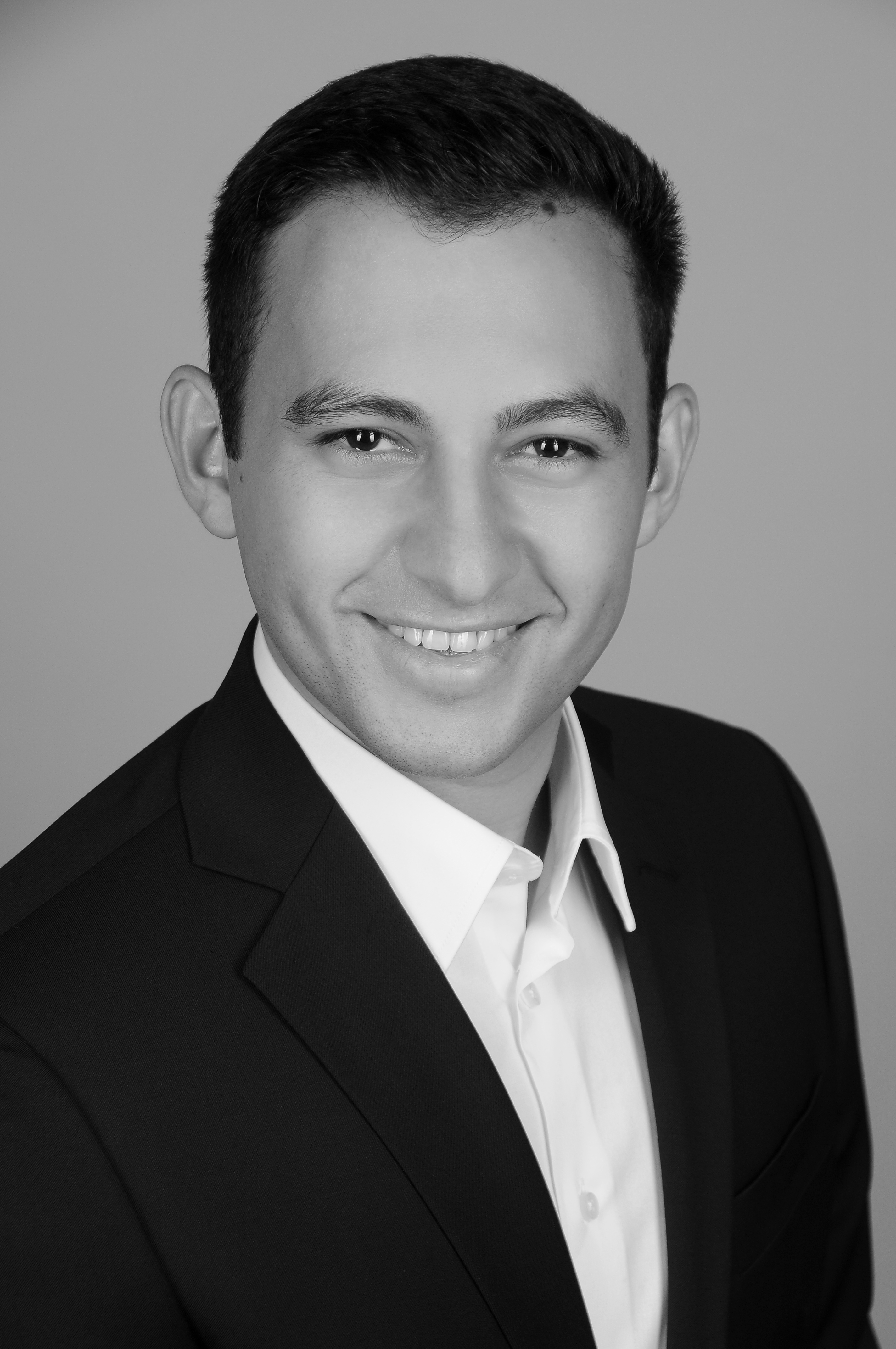}}]{Marco Saur}
	Marco Saur received the MS degree in electrical engineering and information technology from the Technical University of Munich (TUM) in 2016. He is currently working as a software engineer at Kinexon. His research interests include low-power wireless communication, mobile computing and embedded systems.
\end{IEEEbiography}
\vspace*{-1.1cm} 	
\begin{IEEEbiography}[{\includegraphics[width=1in,height=1.25in,clip,keepaspectratio]{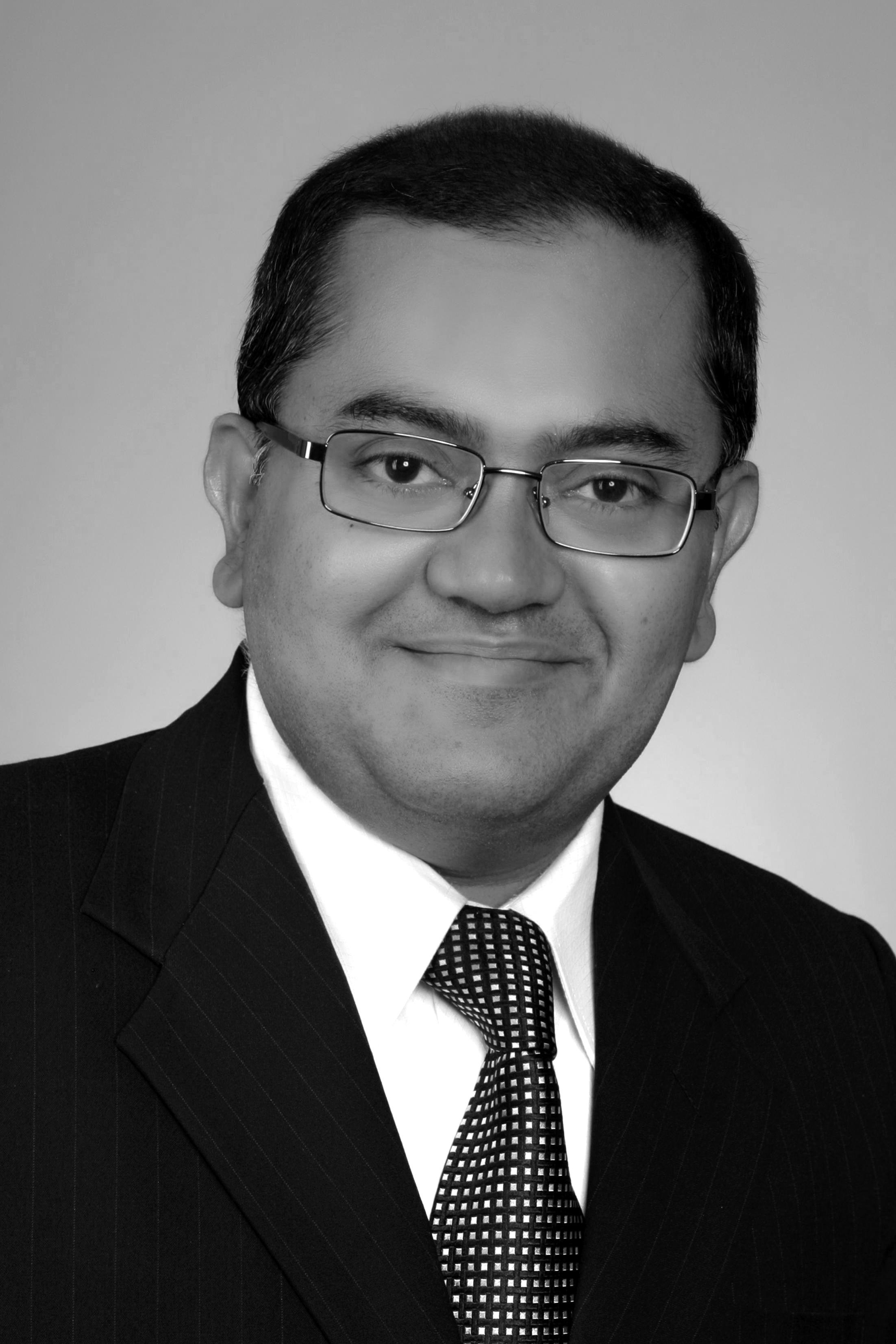}}]{Samarjit Chakraborty}
is a William R. Kenan, Jr. Distinguished Professor in computer science at the University of North Carolina at Chapel Hill, USA. From 2008 to 2019, he was a professor at TU Munich in Germany and an assistant professor at the National University of Singapore from 2003 – 2008. He obtained his Ph.D. in electrical engineering from ETH Zurich in 2003. His research interests include system-level design of embedded and cyber-physical systems, with applications in automotive, healthcare and sensor networks.
\end{IEEEbiography}
\vfill
\pagebreak
\appendix
\input{sections/app_constraints.tex}
\end{document}

%% file: sections/abstract.tex
Neighbor discovery (ND) protocols are used for establishing a first contact between multiple wireless devices. The energy consumption and discovery latency of this procedure are determined by the parametrization of the protocol. In most existing protocols, reception and transmission are temporally coupled. Such schemes are referred to as \textit{slotted}, for which the problem of finding optimized parametrizations has been studied thoroughly in the literature.
However, slotted approaches are not efficient in applications in which new devices join the network gradually and only the joining devices and a master node need to run the ND protocol simultaneously. For example, this is typically the case in IoT scenarios or Bluetooth Low Energy (BLE) piconets.
Here, protocols in which packets are transmitted with periodic intervals (PI) can achieve significantly lower worst-case latencies than slotted ones. For this class of protocols, optimal parameter values remain unknown. To address this, we propose an optimization framework for PI-based BLE-like protocols, which translates any specified duty-cycle (and therefore energy budget) into a set of optimized parameter values. We show that the parametrizations resulting from one variant of our proposed scheme are optimal when one receiver discovers one transmitter, and no other parametrization or ND protocol -- neither slotted nor slotless -- can guarantee lower discovery latencies for a given duty-cycle in this scenario. 
Since the resulting protocol utilizes the channel more aggressively than other ND protocols, beacons will collide more frequently. Hence, due to collisions, the rate of successful discoveries gracefully decreases for larger numbers of devices discovering each other simultaneously. We also propose a scheme for configuring the BLE protocol (and not just BLE-\textit{like} protocols). Though it is not clear whether the resulting parameter values are optimal for BLE, reasonably low worst-case latencies can be guaranteed.

%% file: sections/content.tex
\input{sections/introduction}

\input{sections/related_work}

\input{sections/PI-kM_protocol}
\input{sections/proof_of_optimailty}

\input{sections/implementation}
\input{sections/ble}
\input{sections/evaluation_ii}

\input{sections/concluding_remarks}

%% file: sections/introduction.tex
\section{Introduction}
\label{sec:introduction}
%Low power mobile ad-hoc networks (MANETs), which provide wireless connectivity between multiple mobile devices without the need for stationary, grid-powered infrastructure, are widely used in applications such as location beacons or contact tracking.
In mobile ad-hoc networks (MANETs), all participating devices are battery-powered and hence, energy-saving communication is a crucial requirement. For establishing a first contact between multiple devices, \textit{\ac{ND}} protocols are used, which attempt to achieve low worst-case latencies with low duty-cycles and hence, energy-consumption.

This paper concerns periodic interval (PI)-based protocols, in which transmissions and reception windows are scheduled periodically. The \ac{BLE} protocol builds upon such a procedure for \ac{ND}, but applies multiple extensions to it. We therefore refer to protocols that rely solely on periodic intervals for scheduling reception windows and transmissions as \textit{\ac{BLE}-like}.
For such protocols, we propose a computationally inexpensive optimization framework. Given a target duty-cycle, our framework will help in computing the protocol parameters that, when only a pair of devices discover each other,  outperform the latencies of a large number of well-known ND protocols, e.g., Disco~\cite{dutta:08}, Searchlight~\cite{bakht:12}, U-Connect~\cite{Kandhalu:10} and G-Nihao~\cite{qiu:16}. Further, PI-based protocols configured using this framework outperform difference codes~\cite{meng:14}, which have been proven to provide the lowest-possible latencies among all slotted protocols. In fact, we prove that when one receiver discovers one transmitter, the parametrizations resulting from our framework are optimal for all specified duty-cycles. In particular, no other parametrization will lead to lower worst-case latencies. Further, we show that no other ND protocol can beat the latencies achieved in such a scenario.

We focus on networks with small numbers of devices carrying out the discovery simultaneously, which require different solutions for ND than larger networks. We next describe two different classes of ND protocols (viz., \textit{slotted} and \textit{periodic interval-based}), and give further details afterwards.

\noindent\textbf{Slotted Protocols:} In this class of \ac{ND} protocols, time is subdivided into multiple slots with equal lengths $d_{sl}$.
In each slot, a device can either be in a sleep-mode or in the active mode.
It transmits a beacon with a length of $d_{a}$ time-units at the beginning and/or at the end of an active slot, while listening for incoming packets between them. The discovery procedure is complete once two active slots of two devices overlap in time. 
The primary objective of such slotted \ac{ND} protocols is to identify a pattern of active slots for which discovery is always guaranteed within a certain number of slots.
In contrast, periodic interval (PI)-based protocols temporally decouple reception and transmission, as described next.

%\begin{figure}[t!]
%\centering
%\includegraphics[width=\linewidth]{images/interval_vs_slotted.png}
%%\includegraphics[width=\linewidth]{images/introfig.png}
%\caption{Neighbor discovery protocols. a) Slotted and b) Purely Interval (PI)-based.}
%\label{fig:interval_vs_slotted} 
%\end{figure}

\noindent\textbf{PI-Based Protocols:} 
Many recent protocols, such as Bluetooth Low Energy~(BLE)~\cite{bleSpec50} or ANT/ANT+~\cite{AntSpec:14}, apply a PI-based discovery scheme.
Here, one device, which is called the \textit{advertiser}, periodically sends packets with a period of $T_a$ time-units. 
The transmission duration of a packet $d_a$ is determined by the number of bytes sent. 
The other device, referred to as the \textit{scanner}, periodically switches on its receiver for a duration called the \textit{scan window} $d_s$, with a repetition period called the \textit{scan interval} $T_s$. For two way discovery, every device both transmits packets and listens for incoming ones. Hence, while slotted protocols are characterized by temporally coupling reception and transmission, PI-based ones on the other hand schedule them independently from each other. 

\noindent\textbf{ND in Small vs. Large Networks:} 
Many wireless networks, in particular in the personal area domain, consist of only a few devices being in discovery mode simultaneously. 
For example, in BLE piconets, two devices usually communicate in a synchronized fashion after the \ac{ND} procedure has completed. Here, \ac{ND} is only used for setting up the connection. Even in setups in which additional devices might join during runtime, only the connection master needs to continuously run a \ac{ND} protocol. Hence, the number of devices performing \ac{ND} simultaneously is limited.
Though the total number of wireless devices is likely to increase in the future, networks with few devices being in discovery mode simultaneously will remain important. For example, IoT setups, in which new devices join the network gradually, are expected to grow in importance.
Whereas slotted \ac{ND} protocols, e.g., \cite{dutta:08, Kandhalu:10,bakht:12, meng:14}, can be efficient solutions in large networks with tens or hundreds of participating devices, they perform inefficiently in scenarios with only a few (e.g., up to ten) devices being in discovery mode simultaneously. This is because slotted schemes imply that transmission and reception take place jointly within each slot, which prevents a systematic optimization of the ratio between time spent for transmission and reception. In contrast, PI-based protocols can send the \textit{optimal} rate of beacons for achieving a low worst-case latency and energy consumption. Our solution utilizes this freedom and schedules beacon transmissions more frequently than in other protocols, leading to low latencies. On the other hand, this leads to a higher channel utilization, which can be a multiple of that of popular slotted protocols. As a result, beacons from different devices overlap with each other more frequently, which can lead to a larger number of failed discoveries when the number of devices is increased.
As we show in this paper, for the relevant case of very few devices discovering each other simultaneously, the fraction of failed discoveries remains reasonably low. The collision probability increases with more devices being in discovery mode simultaneously, but the degradation of the rate of successful discoveries is graceful.

\noindent \textbf{Optimizations in PI-Based Protocols:}
This paper proposes an optimization framework for minimizing the worst-case latency guaranteed by a PI-based protocol for a given duty-cycle.
Our proposed techniques are applicable to several \ac{BLE}-like PI-based protocols, as discussed earlier. The specific connection to \ac{BLE} is addressed in Section~\ref{sec:ble}. 
While a significant amount of research has been carried out to analyze and improve slotted protocols, the optimization of PI-based ones has not been sufficiently studied.
This is mainly due to the popular belief that they cannot guarantee deterministic latencies.
Except for trivial cases like $T_a \leq d_s - d_a$ (i.e., the distance between two packets is below the scan window length) or $T_s = d_s$ (i.e., the scanner is always active), the discovery latencies of PI-based protocols have not been well understood, until it was recently shown~\cite{kindt:15} that deterministic discovery latencies can also be obtained using them. However, the only known generic model for PI-based protocols \cite{kindt:15} is given in the form of a recursive computation scheme, and therefore cannot be ``inverted'' to identify beneficial parameter values. In other words, while \cite{kindt:15} shows how worst case latencies can be computed from protocol parameters, it is not possible to use these results to determine parameters for achieving optimal discovery latencies for a given duty cycle (and hence energy budget). Performing an exhaustive search to find optimal parameter values is not practical either, since there are 3 degrees of freedom (viz., $T_a$, $T_s$ and $d_s$), leading to a computationally infeasible procedure. This paper addresses this problem and proposes a framework to translate any given target duty-cycle into a set of optimized parameter values. In particular, $T_a$, $T_s$ and $d_s$ are derived from a target duty-cycle using closed-form equations, which makes exhaustive searches or other computationally expensive procedures unnecessary.

Moreover, while there are a large number publications that compare the performance of various slotted protocols to each other, what is the performance of PI-based protocols, such as BLE? Due to the lack of optimal parametrizations, their performance still remains unclear. This holds true particularly for networks with few devices carrying out ND in parallel, in which protocols such as BLE are frequently used. We in this paper for the first time compare the performance of the BLE discovery procedure to existing ones.
Besides optimal parameter values, a methodology for comparing the performance of optimized slotless and slotted protocols is required. The worst-case latencies of slotted solutions are proportional to the slot length, for which no lower limit is known. This has prevented reasonable comparisons in the past. In this paper, we derive the practical limits of the slot length and hence make such comparisons possible.

\noindent\textbf{Proposed Parametrization Schemes:}
In this paper we propose multiple related parametrization schemes.
First, the \textit{SingleInt} scheme is proposed. It guarantees discovery within $1 \times T_s$ time-units. As we show in Section~\ref{sec:proofOfOptimality}, such parametrizations provide optimal latencies, if one receiver discovers one sender unidirectionally. Based on this, we propose \textit{SingleInt-BLE}, a variant for configuring \ac{BLE}. The constraints of BLE, such as 3-channel operation, lead to considerably larger latencies of SingleInt-BLE compared to SingleInt. Though the latencies guaranteed by SingleInt-BLE are reasonably low, it is not clear whether there are alternative parametrizations for BLE that lead to even lower latencies for a given duty-cycle. 

The \ac{BLE} protocol uses an asymmetric discovery procedure, which allows us to build upon the SingleInt scheme for its parametrization. However, in symmetric, bi-directional discovery scenarios, a large fraction of discoveries would fail when using the SingleInt scheme. We therefore propose the \textit{MultiInt} scheme, which achieves discovery within multiples of $T_s$, thereby lowering the fraction o failed discoveries. To account for non-idealities of the radio hardware, we furthermore propose the \textit{MultiInt-BC} scheme, which is designed to provide a low failure probability on real-world hardware, while achieving latencies that are only slightly larger than the optimal ones guaranteed by the SingleInt scheme. In particular, for the least beneficial duty-cycle we have considered, a near-optimal latency is guaranteed in $\SI{99.8}{\percent}$ of all discovery attempts, while the remaining ones fail. We will study in Section~\ref{sec:ble} that parametrizations based on the MultiInt scheme cannot be used to configure \ac{BLE} radios, since the resulting parameter values would either not lead to bounded latencies, or would not comply with the \ac{BLE} standard.

\noindent\textbf{Contributions and Organization:}
The major contributions of this paper are the following.
\begin{asparaenum}
	\item We propose the first closed-form optimization framework for PI-based protocols in Section~\ref{sec:PIkMProtocol}.
	It is given by a set of equations for translating any specified target duty-cycle $\eta$ into parameter values for $T_a$, $T_s$ and $d_s$.
	\item We show in Section~\ref{sec:proofOfOptimality} that one receiver discovering one transmitter achieves optimal discovery latencies for every joint duty-cycle, if our proposed SingleInt scheme is used.
	\item As in any model, we made some idealistic assumptions for our framework. A real-world hardware has a number of non-idealities, like turnaround times and clock quantization errors. Therefore, we have taken steps to minimize the differences between the platform and the model. Building upon these steps, in Section~\ref{sec:implementation}, we present a real-world implementation of a PI-based protocol parameterized using the MultiInt-BC scheme on a wireless radio.
	\item We show that the MultiInt-BC scheme achieves lower latencies than multiple existing \ac{ND} protocols in Section~\ref{sec:evaluation}.
	\item We show how the values obtained from our optimization framework can be tuned for configuring BLE in Section~\ref{sec:ble}. To the best of our knowledge, we thereby propose the first closed-form parametrization framework to configure the BLE protocol towards low latencies and energy consumption.
	\item A methodology for comparing the performance of slotted protocols to that of PI-based approaches is not available in the literature. The latencies of slotted protocols are proportional to the slot length, for which a lower limit has to be identified for reasonable comparisons. 
	In any \ac{ND} protocol, even in deterministic ones, every discovery attempt can fail with a certain probability, and two devices never discover each other in such cases. For slotted protocols, we identify the slot lengths that lead to the same failure rates as those of our proposed solution and compare the performance of optimized PI-based protocols with existing approaches in Section~\ref{sec:evaluation}. In a scenario in which two devices discover each other, the MultiInt-BC scheme outperforms all considered slotted \ac{ND} protocols. E.g., when allowing a rate of failed discoveries of $\SI{0.19}{\percent}$ for two devices, difference codes \cite{meng:14}, which achieve the theoretically optimal latencies for slotted protocols, provide by $385 \times$ larger worst-case latencies than our solution on the average of all considered duty-cycles. Further, using experimental measurements on $560,000$ discovery procedures, we demonstrate that the anticipated performance is reached in practice. 
\end{asparaenum}
In the next section, we present related work. Finally, we discuss the implications of our theory in Section~\ref{sec:conclusion}.

%% file: sections/related_work.tex
\section{Related Work}
\label{sec:related_work}

In this section, we give an overview on related work on ND.

\noindent\textbf{Slotted Discovery:}
In slotted ND protocols, time is subdivided into equal-length slots. There are sleep slots and active slots. In each active slot, both transmission and reception take place jointly. Every slotted protocol defines a unique discovery schedule, which determines the set of active slots in each period.
Known deterministic schedules are based on e.g., coprimal intervals \cite{dutta:08}, systematic probing \cite{bakht:12} or cyclic difference sets \cite{meng:14, choi:11}. Some recently proposed protocols, e.g., U-Connect~\cite{Kandhalu:10} and Nihao~\cite{qiu:16}, define dedicated receive and transmit slots, while still using the assumption of slotted time for simplifying the analysis. 
E.g., in Nihao~\cite{qiu:16}, one beacon is sent at the beginning of each transmit slot and the device goes back to a sleep mode afterwards for the rest of the slot length.

While the pattern of active and passive slots has been studied extensively in the literature, the slot size, which also affects the discovery latency, has not been sufficiently studied. In \cite{jin:18}, shrinking the slot length in slots similar to what is used in Disco~\cite{dutta:08} has been considered. It has been concluded that shorter slots lead to a higher rate of failed discoveries. In this paper, we study shrinking the slot length in different slot designs, such that they all achieve the same failure rate.

\noindent\textbf{PI-Based Discovery:}
As already mentioned, in this class of protocols, reception and transmission are scheduled using periodic intervals. Until recently, the behavior of PI-based protocols has not been well understood. With BLE applying a PI-based scheme, its discovery latencies received attention by the community and a comprehensive latency model \cite{kindt:15} has been proposed recently. However, how optimal parametrizations can be computed has not been addressed so far.
An existing approach for parameter optimization \cite{kindt:17a} concerns achieving low mean latencies by exploiting mutual assistance. It considers modified PI-based schemes, where a device that receives a packet schedules an additional one, based on a received hint on the next reception window of its opposite device.
Another important recent approach for parameter optimization is BLEND~\cite{julien:17}, which considers optimizing similar parametrizations as the \textit{SingleInt} scheme proposed in this paper. The main differences are as follows. BLEND relies on exhaustive searches for two out of three parameters, i.e., an algorithm iterates through all possible configurations and identifies the best one. Hence, due to the computational complexity, the parameter values need to be computed offline. Our proposed technique further improves the BLEND approach by giving optimal parametrizations using closed-form equations. Another difference is that the BLEND optimization algorithm accounts for additional optimization goals, such as different collision probabilities among multiple nodes, whereas our optimization always minimizes the discovery latency for a pair of devices, without controlling the corresponding collision probability. Further, BLEND optimizes parameters in conjunction with the random delay applied in BLE, whereas we mainly study discovery without any randomization. Since BLEND applies an exhaustive parameter search, the resulting intervals are potentially also near-optimal. However, the discretized search space (e.g., a discretization of $\SI{1}{ms}$ was assumed in ~\cite{julien:17}) might lead to small deviations from the theoretically optimal values.
In summary, no comparable, closed-form framework for optimizing the parameters of PI-based protocols towards low discovery latencies exists.\\
\noindent\textbf{Other Approaches:}
As an alternative to duty-cycling the main radio, wake-up receivers can be used~\cite{loreti:19}. Further, special \ac{ND} solutions can be applied when the clocks of the participating devices are already synchronized, e.g., using GPS~\cite{li:14}. While most protocols assume that the designer manually chooses the duty-cycle, \cite{bracciale:16} concerns finding the optimal duty-cycle for a given environment. Further,~\cite{yang:15} concerns minimizing the probability of failed discoveries.\\
\noindent\textbf{Fundamental Limits of Neighbor Discovery:}
A fundamental performance limit of slotted ND protocols has been known since 2003. It has been shown in~\cite{zheng:03} that no slotted ND protocol can guarantee discovery within $T$ slots by having less than $\sqrt{T}$ active slots per $T$. However, this bound is only valid for slotted protocols, in which reception and transmission are temporally coupled. For increasingly popular slotless protocols such as BLE, this bound does not apply. Further, it only gives a relation between the number of active slots and the discovery latency in terms of slot lengths. The discovery latency in terms of time is proportional to the slot length, and no lower limit of this length has been known.
Recently, in \cite{kindt:19}, a lower bound on the discovery latency that any neighbor discovery protocol could guarantee for a certain duty-cycle has been derived. However, the results from \cite{kindt:19} do not translate into the design of a protocol that actually utilizes this bound. In contrast, the goal of this paper is optimally parametrizing BLE-like protocols.

%% file: sections/PI-kM_protocol.tex
\section{Optimization Framework}
\label{sec:PIkMProtocol}

In this section, we describe our proposed framework for parametrizing PI-based protocols.
As already mentioned, in PI-based protocols, beacons are transmitted with an interval $T_a$, whereas reception windows of length $d_s$ are scheduled periodically every $T_s$ time-units. Each beacon transmission takes $d_a$ time-units.
Our objective is to derive a set of equations that translate any specified target duty-cycle $\eta$ into parameter values for $T_a$, $T_s$ and $d_s$, which lead to minimal latencies. 

In this section, we consider a pair of one sender and one receiver with a sum of duty-cycles $\eta$. We extend our results towards symmetric (i.e., both devices receive and transmit), bi-directional \ac{ND} in Section~\ref{sec:implementation} and \ac{BLE} in Section~\ref{sec:ble}.

\subsection{Performance Metrics}
We now define the major properties and performance metrics of PI-based \ac{ND} protocols.

The \textit{duty-cycle} $\eta$ is the main energy metric of a \ac{ND} protocol. It corresponds to the fraction of time a device is active. For PI-based protocols, it is defined as
\begin{equation}
\label{eq:etaDef}
\eta = \frac{d_s}{T_s} + \alpha \cdot \frac{d_a}{T_a} = \frac{T_a d_s + \alpha T_s d_a}{T_a T_s}.
\end{equation}
Here, $\alpha$ is a balancing factor to account for different energy consumption for transmission and reception. 
For the simplicity of exposition, we assume $\alpha=1$ for the rest of this paper, but all of our results will easily extend to other values, too.

Let two devices come into their range of reception at a certain point in time $t_0$.  At $t_0$, the first packet of the sender has a temporal offset of $\Phi_0$ time-units from its preceding scan window of the receiver (cf. Figure~\ref{fig:simpleLatencyComputationModel}a). Since there is no previous synchronization between any two devices, $\Phi_0$ has a random value between $0$ and $T_s$ time-units. 
The discovery procedure is completed, once a beacon of the sender entirely coincides with a reception window of the receiver. 
Hence, the effective length of a scan window is $d_s - d_a$ time-units, since a beacon transmission starting within the last $d_a$ time-units of a scan window will only partially overlap with it and hence cannot be received successfully. 
%Receivers that can extend their scan window upon a partial reception are also feasible. In such cases, the effective scan window length would be $d_s$. 

The \textit{worst-case latency} $d_m$ is given by the time measured from $t_0$ (i.e., from the point in time at which both devices come into range) until a beacon entirely overlaps with a scan window for \textit{every} initial offset $\Phi_0$. 
After both devices have been brought into range, some time might pass until the next beacon is sent.
The lowest latency measured from the first beacon that is sent in range until the discovery procedure is successful is called the  \textit{packet-to-packet discovery latency} $d_m^*$.
The latency $d_m$ is relevant when two devices that are already running are brought into range. If the devices are already in range but the sender is switched on at some point in time and immediately starts transmitting while the receiver is already running, $d_m^*$ is relevant. This is for example the case when a gadget needs to be connected to a smartphone. Here, the smartphone will, in most cases, already be scanning, while the gadget is suddenly switched on. 
Since beacons are sent with periodic intervals and since two devices might come into range by up to $T_a$ time-units before the first beacon in range is sent, it is $d_m = d_m^* + T_a$. Clearly, it is always $d_m^* < d_m$. In this paper, we concern optimizing $d_m$.

\subsection{Overview}
A large set of values for $T_a$, $T_s$ and $d_s$ that realize the same duty-cycle exist, of which multiple values potentially optimize $d_m$.
In our proposed optimization, we first pre-constrain the set of possible values by establishing multiple relations between these parameters. These relations mainly exploit the following insight. For any given offset $\Phi_0$, there should be exactly one beacon within $d_m$ that falls into exactly one scan window of the receiver. If there were two successful beacons sent within $d_m$, one of them would be redundant, i.e., omitting this beacon would not increase the worst-case latency $d_m$, but reduce the duty-cycle. More details on the generic properties that any \ac{ND} protocol needs to fulfill for performing optimally can be found in \cite{kindt:19}.

We next systematically optimize the parameter values that lie within the set of constrained values. Though the set of parameters that are excluded by these constraints might also contain optimal parameter values, we formally prove in Section~\ref{sec:proofOfOptimality} that for every $\eta$, the parameter values computed by our framework always lead to the theoretically minimal worst-case latencies $d_m$.

We consider three different cases. Each of them pre-constrains the design space to a different set of values. As will become clear in Section~\ref{sec:implementation}, all of them perform similarly for the one-way case (i.e., one sender and one receiver). However, one of them is beneficial for being applied to symmetric \ac{ND} scenarios, whereas another one is well-suited for parametrizing BLE. The schedules of beacon transmissions and reception windows for these cases are shown in Figure~\ref{fig:simpleLatencyComputationModel}. The ruled boxes depict the scan windows, whereas the hatched vertical bars show the beacon transmissions. 
First, we consider the situation with $T_a \leq d_s - d_a$, as depicted in Figure~\ref{fig:simpleLatencyComputationModel}a).
\subsection{Optimization Scheme a): SingleInt}
\label{sec:PI_0M}
\begin{figure*}[t!]
\centering
\includegraphics[width=\linewidth]{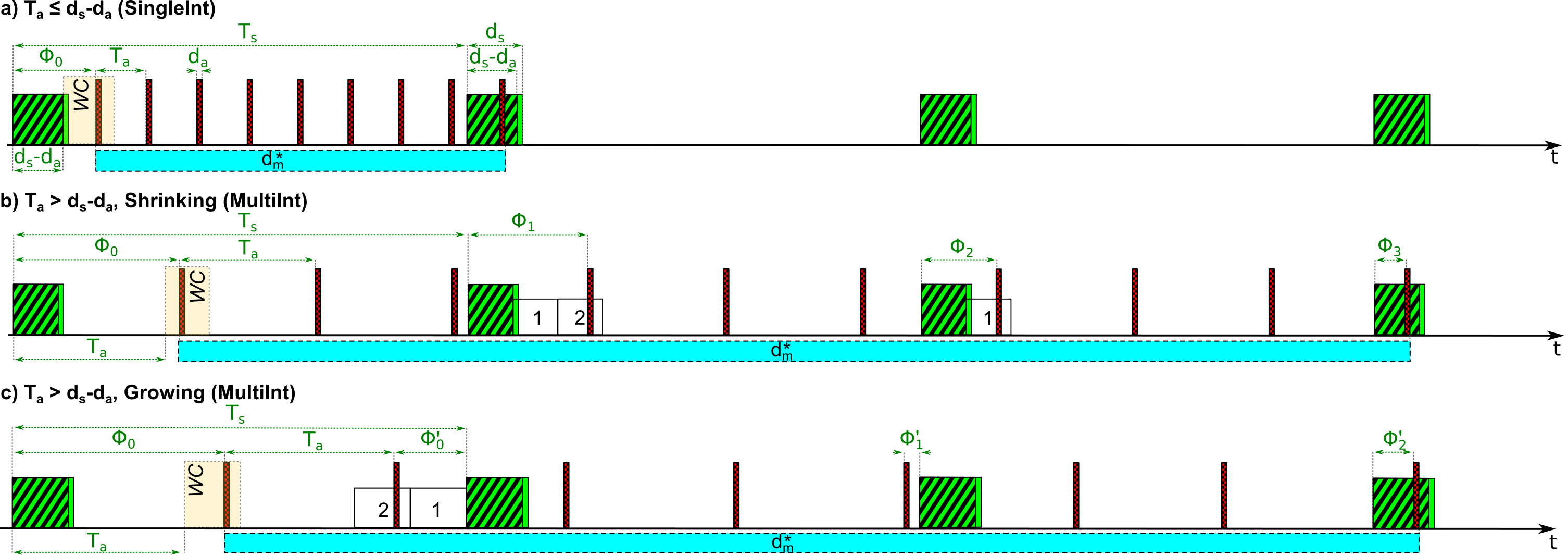}
\caption{Model for a) $T_a \leq d_s - d_a$ and b), c) $T_a > d_s - d_a$. Vertical bars depict beacons, hatched rectangles scan windows. The rectangles tagged with ``\textit{WC}'' depict the ranges of offsets $\Phi_0$ for which the worst-case occurs.}
\label{fig:simpleLatencyComputationModel} 
\end{figure*}

Figure~\ref{fig:simpleLatencyComputationModel}a) shows the beacon flow for a situation with $T_a \leq d_s - d_a$. We call this parametrization scheme \textit{SingleInt}, since it guarantees discovery within one instance of the interval $T_s$. Here, the temporal distance between any two beacons is no more than $d_s - d_a$ time-units, and hence, there is no scan window into which no beacon falls.
Recall that discovery is successful once a beacon is sent entirely within a reception window of the receiver. The worst-case occurs e.g., for values of $\Phi_0$ that are infinitesimally larger than $d_s - d_a$, as depicted in Figure~\ref{fig:simpleLatencyComputationModel}a). The packet-to-packet worst-case latency $d_m^*$, which is indicated by the hatched bar under the time-axis in Figure~\ref{fig:simpleLatencyComputationModel}a), is given by:  

\begin{equation}
\label{eq:dMaxPureOrder0}
d_m^* = \underbrace{\left\lceil \frac{T_s - (d_s - d_a)}{T_a}\right\rceil}_{\mathcal{A}} \cdot T_a + \underbrace{d_a}_{\mathcal{B}}
\end{equation}
Equation~\ref{eq:dMaxPureOrder0} can be explained as follows. 
The distance between the end of one scan window and the beginning of the subsequent one is $T_s - d_s$ time-units. Recall that a beacon transmission beginning within the last $d_a$ time-units of a scan window is not received successfully, and hence the effective length of a scan window is $d_s - d_a$. Therefore, this distance is effectively increased to $T_s - (d_s - d_a)$ time-units. The number of advertising intervals that at least partially ``fit'' into this distance is given by the Term $\mathcal{A}$. In the worst case, after this number of advertising intervals, a beacon overlaps with a scan window and hence, discovery is guaranteed. Term $\mathcal{B}$ accounts for the transmission duration of the successful beacon. 
%Recall that $d_m = d_m^* + T_a$, since both devices might come into range by up to $T_a$ time-units before the first beacon is sent. We next attempt to minimize $d_m$.

\subsubsection{Optimizing $T_s$}
\label{subsubsec:optimizingTs}
Our goal is to identify those values of $T_a$, $T_s$ and $d_s$, that lead to the lowest worst-case discovery latency $d_m$ for every given duty-cycle. 
Equation~\ref{eq:dMaxPureOrder0} implies that $d_m = d_m^* + T_a$ is equal for all values of $T_s$, for which the ceiling function in Term $\mathcal{A}$ does not change its value. Hence, in each such constant segment of Term $\mathcal{A}$, we should chose the largest possible value of $T_s$. In other words, the function $d_m(T_s)$ grows in multiple steps caused by the ceiling-function in Term $\mathcal{A}$ wrapping around to its next higher value, and optimal values of $T_s$ are those that directly precede such steps. Smaller values of $T_s$ would increase the duty-cycle spent for reception (i.e., $\nicefrac{d_s}{T_s})$. Recall from Equation~\ref{eq:etaDef} that if a certain duty-cycle $\eta$ is specified, a larger value of $T_s$ will allow for a smaller value of $T_a$, while still realizing the given~$\eta$. A smaller value of $T_a$ will, as long as Term $\mathcal{A}$ in Equation~\ref{eq:dMaxPureOrder0} does not wrap around (which we have excluded by our choice of $T_s$), minimize $d_m$.
For obtaining those values of $T_s$ that directly precede a step of $d_m(T_s)$, we have to satisfy the following.
\begin{equation}
\label{eq:choosingTsRelation}
\frac{T_s - (d_s - d_a)}{T_a} = M,\ M = 1,2,3,...
\end{equation}
From Equation~\ref{eq:choosingTsRelation} follows that
\begin{equation}
\label{eq:choosingTs_PI0M}
T_s = M \cdot T_a + d_s - d_a.
\end{equation}

\subsubsection{Optimizing ${T_a}$ and ${d_s}$}
\label{subsubsec:optimizingTa}
Inserting Equation~\ref{eq:choosingTs_PI0M} into Equation~\ref{eq:dMaxPureOrder0} and setting $d_m = d_m^* + T_a$ leads to the following latency.
\begin{equation}
\label{eq:0MdmSimplified}
d_m = (M+1) \cdot T_a + d_a
\end{equation}
For a given value of $T_s$ and $\eta$, according to Equation~\ref{eq:etaDef}, $T_a$ is a function of $d_s$. Therefore,  
for analyzing which value of $d_s$ minimizes $d_m$, we cannot directly form the derivative of Equation~\ref{eq:0MdmSimplified} by $d_s$.
We first solve Equation~\ref{eq:etaDef} by $T_a$ and insert the result into Equation~\ref{eq:0MdmSimplified}.
This gives us a relation between $d_m$ and the specified duty-cycle $\eta$, which still depends on the unknown values $d_s$ and $M$.

One can show that the derivative $\frac{\mathrm{d}d_m}{\mathrm{d}d_s}$ is positive for all values of $\eta < 1$. 
Since it represents the slope of $d_m$, we can conclude that the smallest possible value of $d_s$ minimizes $d_m$. 
Because we require $T_a \leq d_s - d_a$ for Case a) (smaller values of $d_s$ would lead to a different scenario, since the discovery would not always complete within one instance of $T_s$, cf. Figure~\ref{fig:simpleLatencyComputationModel}a)), the smallest possible value we can assign to $d_s$ is $T_a + d_a$. 
Hence, we can establish the following relation between $d_s$ and $T_a$.
\begin{equation}
\label{eq:0M-Ta}
T_a = d_s - d_a
\end{equation}
%Figure~\ref{fig:PI0MTaValues} annotates the values of $T_a$ given by Equation~\ref{eq:0M-Ta} and their corresponding latencies within the whole design space for a given value of $d_s$. As can be seen, the design space consists of a large number of irregular peaks, which are avoided by the parameters chosen. 

Using Equation~\ref{eq:0M-Ta}, Equation~\ref{eq:choosingTs_PI0M} can be expanded to the following term.
\begin{equation}
\label{eq:choosingTs_expanded}
T_s = (M+1) \cdot (d_s - d_a)
\end{equation}
Further, inserting $T_a$ from Equation~\ref{eq:0M-Ta} into Equation~\ref{eq:etaDef} and solving this by $d_s$ leads to the following equation.
\begin{equation}
\label{eq:0M-dsByEta}
d_s = \frac{(M+1)\cdot(1 + \eta)\cdot d_a}{\eta \cdot (M+1) - 1}
\end{equation}
%\begin{figure}[t!]
%\centering
%\includegraphics[width=\linewidth]{images/sweep_0m_kmp_kmm.png}
%\caption{Discovery latencies for different values of $T_a$ and $T_s$ and valuations given by $T_a = d_s - d_a$ (i.e., SingleInt) and by $T_a =  %\nicefrac{1}{k} \cdot (T_s + d_s - d_a)$ for different values of $k$.}
%\label{fig:PI0MTaValues} 
%\end{figure}
We now established a set of closed-form equations for $T_a$, $T_s$ and $d_s$. The only undetermined value in these equations is $M$, which we optimize next. 

\subsubsection{Optimizing $M$}
\label{subsubsec:optimizingM}
From Equation~\ref{eq:0M-Ta} and Equation~\ref{eq:0MdmSimplified} follows that $d_m = (M+1) \cdot (d_s - d_a) + d_a$. 
By inserting $d_s(\eta)$ from Equation~\ref{eq:0M-dsByEta}, we obtain the following latency.
\begin{equation}
\label{eq:dmExpanded}
d_m = (M+1)\cdot \left(\frac{d_a(M+1)(\eta + 1)}{\eta(M+1) - 1} - d_a\right) + d_a
\end{equation}
By differentiating $d_m$ from Equation~\ref{eq:dmExpanded} (which we attempt to minimize) and requiring $\frac{d d_m}{d M} = 0$, one can identify a local latency minimum at:

\begin{equation}
\label{eq:pi0MMopt}
M_{opt} = \frac{\sqrt{1+\eta} + 1}{\eta} - 1
\end{equation}
Using this, $T_a$, $T_s$ and $d_s$ are given by closed-form equations that only depend on the target duty-cycle $\eta$ that needs to be realized. 
As we show in Section~\ref{sec:proofOfOptimality}, the resulting values are optimal, and no other values lead to lower worst-case latencies.

\subsubsection{Constraints}
\label{sec:PI0M_constraints}
Note that the transmission duration of a beacon, $d_a$, cannot be optimized, since it is given by the hardware or by the application, which might require a certain number of bytes to be sent in each packet. 
Furthermore, while radios are capable of realizing different lengths of the scan window $d_s$, real-world hardware imposes a certain lower limit $d_{s,m}$.
When expanding the inequality $d_{s}(\eta) \geq d_{s,m}$ using Equation~\ref{eq:0M-dsByEta}, we obtain
\begin{equation}
\label{eq:0M_constraint_inequality}
d_s(\eta) = \frac{(M+1)\cdot(\eta + 1)\cdot d_a}{\eta \cdot (M+1) - 1} \geq d_{s,m}.
\end{equation}
From the requirement that $d_s(\eta) > 0$ follows that the denominator in Equation~\ref{eq:0M_constraint_inequality} must be positive, and hence it is
\begin{equation}
\label{eq:pi0MEtaMin}
M > \frac{1}{\eta} - 1 = M_{min}.
\end{equation}
Using this, we can reformulate Equation~\ref{eq:0M_constraint_inequality} as follows.
\begin{equation}
(M+1) \cdot (\eta + 1) \cdot d_a \geq d_{s,m} \cdot (\eta \cdot (M+1) - 1) 
\end{equation} 
When solving this by $M$ (while accounting for potentially different signs of a double-sided multiplication), 
we obtain that if $\eta > \frac{d_a}{d_{s,m} - d_a}$, $M \leq M_{max}$, where 
\begin{equation}
\label{eq:pi0MEtaMax}
M_{max} = \frac{d_{s,m} (\eta - 1) - d_a (\eta + 1)}{d_a ( \eta + 1) - \eta d_{s,m}}.
\end{equation}
If $\eta \leq \frac{d_a}{d_{s,m} - d_a}$, no constraints on $M$ apply.
Considering these constraints, we always set $M$ to $\mbox{round}(M_{opt})$, or to the value being closest to $\mbox{round}(M_{opt})$ that lies within $[\lceil M_{min} \rceil, \lfloor M_{max} \rfloor]$, respectively.
Since $\lceil M_{min} \rceil \leq \lfloor M_{max} \rfloor$, a conservative bound on the maximum duty-cycle that can be realized using this scheme is obtained from solving $M_{min} + 1 \leq M_{max} - 1$, which leads to
\begin{equation}
\label{eq:pi0metamax}
\eta_{max} = \frac{3 d_a + \sqrt{d_a (d_a + 8 d_{s,m})}}{4 (d_{s,m} - d_a)}.
\end{equation}
A step-by-step derivation of these constraints is given in Appendix~\ref{sec:app_constraints}.
In addition to these constraints, $T_a$ always needs to exceed $d_a$ and $T_s$ needs to exceed $d_s$. This is intrinsically fulfilled by the equations presented above. The resulting values meet all requirements for using them on state-of-the-art hardware, as we show in Section~\ref{sec:implementation}.

As will become clear in Section~\ref{sec:implementation}, the SingleInt scheme is beneficial for one-way discovery, where one sender discovers one receiver. In contrast, the schemes described next will be more beneficial for parametrizing PI-based protocols in symmetric, two-way scenarios.

\subsection{Schemes b) and c): MultiInt}
We now consider values with $T_a > d_s - d_a$ and study two possible parametrization Schemes~b) and c). 
These schemes will guarantee discovery within a certain multiple of $T_s$.
For Scheme~b), let $k$ be the largest multiple of $T_a$, such that $k \cdot T_a < T_s$ and $T_s - k_f \cdot T_a < \nicefrac{1}{2} \cdot T_a$. This is given by $k_f = \lfloor \nicefrac{T_s}{T_a}\rfloor$. For example, in Figure~\ref{fig:simpleLatencyComputationModel}b), $k_f = 3$.

Let us now consider the offset $\Phi_1$ between the second scan window and the beacon that directly succeeds it. 
Since the time $k_f \cdot T_a$ is slightly (i.e, by less than $\nicefrac{1}{2} \cdot T_a$ time-units) shorter than $T_s$, the offset $\Phi_2$ between the third scan window and the beacon that directly succeeds it is given by (cf. Figure~\ref{fig:simpleLatencyComputationModel}b))
\begin{equation}
\label{eq:gammaShrinkage}
\Phi_2 = \Phi_1 + k_f \cdot{T_a} - T_s = \Phi_1 - \gamma,
\end{equation}
with the shrinkage $\gamma = T_s - k_f \cdot T_a$.
We can generalize this towards the other offsets $\Phi_i$, $i = 1,2,...$ as follows. Starting from any value of $\Phi_1 > 0$, after every instance $i$ of the scan interval $T_s$, the offset $\Phi_i$ between a scan window and the next beacon that is sent after its beginning is reduced by $\gamma$ time-units, and it is
\begin{equation}
\label{eq:gammaShrinkageGeneral}
\Phi_i - \Phi_{i+1} = \gamma,\ i = 1,2,...
\end{equation}
Clearly, if $\gamma < d_s - d_a$ (i.e., the shrinkage per $T_s$ does not exceed the effective scan window length), after a finite number of interval instances, the remaining offset $\Phi_i$ will fall below $d_s - d_a$, and hence a beacon will coincide with a reception window.
For example, in Figure~\ref{fig:simpleLatencyComputationModel}b), $\Phi_3$ is the first offset that is smaller than $d_s - d_a$, and hence, a beacon is received by the last depicted scan window. Since the temporal distance between subsequent scan windows and their neighboring beacons shrinks successively, Scheme~b) is referred to as \textit{shrinking}.

For computing the worst-case latency, consider the beacon that directly succeeds the second scan window in Figure~\ref{fig:simpleLatencyComputationModel}b) (i.e., the beacon corresponding to $\Phi_1$). We first describe the computation of the latency starting from this beacon, and then describe the additional latency  induced for reaching this beacon from the first one in range. 

Starting from the beacon succeeding the second scan window, a later beacon will overlap with a scan window after the offset has been reduced by a certain number of steps of length $\gamma$. For example, in the figure, two ranges of offsets are marked with the numbers \textit{1} and \textit{2}, indicating that starting from a beacon sent within these ranges of offsets, 1 or 2 steps of $\gamma$ are needed until a scan window is reached. The maximum value of $\Phi_1$ is $T_a$ time-units (since for larger offsets, an earlier beacon would be sent after the beginning of the second scan window and hence define $\Phi_1$). Therefore, the maximum distance between the end of the second scan window and its succeeding beacon is $T_a - (d_s - d_a)$ time-units. Measured from the first beacon that is sent after the second scan window, the worst-case latency $d_{m,p}^*$ would therefore be as follows.
\begin{equation}
\label{eq:dm_partial_k1}	
d_{m,p}^* = \left\lceil \frac{T_a - (d_s - d_a)}{\gamma} \right\rceil \cdot \left\lfloor \frac{T_s}{T_a} \right\rfloor T_a + d_a
\end{equation}
The first ceiling term accounts for the number of $\gamma$-steps until a beacon falls into the reception window, whereas the term
$\left\lfloor \nicefrac{T_s}{T_a} \right\rfloor T_a$ accounts for the number of advertising intervals per step of $\gamma$. Finally, $d_a$ is the transmission duration of the successful beacon.

We yet have to account for the latency that is induced before sending the earliest beacon that succeeds the second scan window (i.e., the beacon that corresponds to $\Phi_1$). Depending on the value of $\Phi_0$, multiple advertising intervals need to be added to the latency from Equation~\ref{eq:dm_partial_k1} to account for this.
We have to distinguish between the following two ranges of $\Phi_0$.
\begin{asparaenum}
\item $\Phi_0 \leq T_a$: 
When $\Phi_0 \leq T_a$, then $\Phi_1 \leq T_a - \gamma$ (cf. Equation~\ref{eq:gammaShrinkage}). 
Therefore, though $\Phi_0 \leq T_a$ leads to the maximum number of $T_a$ intervals until reaching the first beacon sent after the second scan window, the number of $\gamma$-steps needed until a beacon actually coincides with a scan window is reduced by 1 compared to the worst-case. Hence, for computing the overall worst-case latency, we do not need to consider this range of offsets.
\item $\Phi_0 > T_a$: For the maximum number of $\gamma$-steps, $\Phi_1$ needs to lie within $]T_a - \gamma, T_a$] (where ``$]$'' means that the left interval border is not included).
Because of Equation~\ref{eq:gammaShrinkageGeneral}, the corresponding range of $\Phi_0$ that leads to this range of $\Phi_1$ and maximizes the amount of time from the first beacon in range until reaching the first beacon after the second scan window is $]T_a, T_a + \gamma$], as highlighted by the frame tagged with ``\textit{WC}'' in Figure~\ref{fig:simpleLatencyComputationModel}b).
This amount of time is therefore $T_s + \Phi_1 - \Phi_0 = T_s - \gamma$. By inserting $\gamma = T_s - k_f \cdot T_a$, it becomes $k_f \cdot T_a$. Therefore, to account for the time that passes from the first beacon in range to the first beacon sent after the second scan window, we have to add $\left \lfloor \nicefrac{T_s}{T_a}\right\rfloor \cdot T_a$ time-units to the latency given by Equation~\ref{eq:dm_partial_k1}.
\end{asparaenum}
Considering this, the worst-case discovery latency for Case~b) is as follows.
\begin{equation}
\label{eq:dMaxPureOrder1s}
d_m = \underbrace{\left\lfloor \frac{T_s}{T_a} \right\rfloor T_a}_{\mathcal{A}} + \underbrace{\left\lceil \frac{T_a - (d_s - d_a)}{\gamma} \right\rceil \cdot \left\lfloor \frac{T_s}{T_a} \right\rfloor T_a}_{\mathcal{B}}  + \underbrace{T_a}_{\mathcal{C}} +\  d_a
\end{equation}
Here, Term $\mathcal{A}$ accounts for the latency induced by reaching the first beacon sent after the second scan window, whereas Term $\mathcal{B}$ accounts for the latency until reaching a scan window from this bacon. Recall that $d_m = d_m^* + T_a$, which Term $\mathcal{C}$ accounts for.

In Case~c), let $k_c$ be the smallest multiple of $T_a$, such that $k_c \cdot T_a > T_s$ and $k_c \cdot T_a - T_s < \nicefrac{1}{2} \cdot T_a$, i.e., $k_c = \lceil \nicefrac{T_s}{T_a} \rceil$. For example, in Figure~\ref{fig:simpleLatencyComputationModel}c), $k_c = 3$.
Because $\Phi_i$ grows after every instance $i$ of $T_s$, this case is referred to as \textit{growing}.
Here, we have to consider the offsets $\Phi_i^\prime$ between the closest beacon that precedes a scan window and the beginning of this scan window (in contrast to $\Phi_i$, which is measured between a scan window and its succeeding beacon). For example, in Figure~\ref{fig:simpleLatencyComputationModel}c), the initial offset $\Phi_0$ implies a certain offset $\Phi_0^\prime$ between the second beacon and the second scan window. Such offsets shrink after every instance of $T_s$. Similarly to Case b), it is $\Phi_{i}^\prime - \Phi_{i+1}^\prime = \gamma$, with $\gamma = k_c \cdot T_a - T_s$. Again, we have to chose $T_a$ and $T_s$, such that $\gamma \leq d_s - d_a$. The deduction of the worst-case latency $d_m$ works similarly to Case b). For Case~c), the worst-case latency is

\begin{equation}
	\label{eq:dMaxPureOrder1g}
	d_m = \underbrace{\left\lceil \frac{T_s}{T_a}\right\rceil T_a}_{\mathcal{A + C}} + \underbrace{\left\lceil \frac{T_a - (d_s - d_a)}{\gamma} \right\rceil \cdot \left\lceil \frac{T_s}{T_a} \right\rceil T_a}_{\mathcal{B}} + d_a.
\end{equation}
We next optimize $T_a$, $T_s$ and $d_s$ for minimizing the latencies in Case~b) and c).
\subsubsection{Choosing ${T_a}$}
Recall that the worst-case latency for Case~b), as given by Equation~\ref{eq:dMaxPureOrder1s}, is composed of three parts $\mathcal{A}$, $\mathcal{B}$ and $\mathcal{C}$. Starting from the first beacon that is sent after the second scan window, the distance between neighboring pairs of scan windows and beacons is reduced by multiple steps of $\gamma$, which Term $\mathcal{B}$ accounts for.
It follows from Equation~\ref{eq:dMaxPureOrder1s} that every additional step of length $\gamma$ induces a latency that already reaches that of Term $\mathcal{A}$. Hence, the optimal parametrization will minimize the number of $\gamma$-steps. The same also holds true for Case~c). 
For this reason, we have to maximize the value of $\gamma$. The maximum possible value of $\gamma$ is $d_s - d_a$, since larger values would lead to offset shrinkages or growths per scan interval of $\gamma > d_s - d_a$, and hence the scan window would be missed for some offsets $\Phi_0$. Note that if $\gamma$ was smaller than $d_s - d_a$, for some offsets, two beacons would overlap with two different scan windows within $d_m$. Since one of these beacons would be redundant, i.e., it could be removed without increasing $d_m$, such parametrizations are not optimal.

From $\gamma = | k_c \cdot T_a - T_s|$,  it follows that $\gamma = d_s - d_a$ can be realized by setting $k \cdot T_a$ by $d_s - d_a$ time-units shorter ($k= k_f$) or longer ($k = k_c$) than one scan interval:
\begin{equation}
\label{eq:choosingTa}
k \cdot T_a = T_s \pm (d_s - d_a)
\end{equation}

Both the ``$+$''-operation and the ``$-$''-operation in Equation~\eqref{eq:choosingTa} lead to potentially optimal parametrizations. We in the following restrict ourselfs to $k_c \cdot T_a = T_s + (d_s - d_a)$, since this case is sufficient for obtaining an optimal parametrization for every duty-cycle. The analysis for the other case works similarly.

Because $k_c \cdot T_a > T_s$ and $k_c \cdot T_a - T_s < \nicefrac{1}{2} \cdot T_a$, such configurations always lead to Case~c)
%\footnote{For $M = 1$, the distances between a scan window and its preceding beacon and between a scan window and its succeeding beacon are both reduced in steps of length $\gamma$. Despite of this, the worst-case latency nevertheless corresponds to that of Case~c).} 
(cf. definition of Case~c)). Hence, we in the following only need to consider the latency given by Equation~\ref{eq:dMaxPureOrder1g}. %Figure~\ref{fig:PI0MTaValues} shows the interval lengths and corresponding latencies that correspond to $k \cdot T_a = T_s + (d_s - d_a)$ for different values of $k$ and a fixed value of $d_s$. As can be seen, such parametrizations avoid the numerous latency peaks in the design space.

\subsubsection{Relation between ${T_s}$ and ${d_s}$}
%\begin{figure}[t!]
%\centering
%\includegraphics[width=\linewidth]{images/multipleM.png}
%\caption{Worst-case latency $d_m^*$ for $T_s = \SI{5}{s}$ and $T_a = \nicefrac{1}{k} \cdot (d_s + d_s - d_a)$, $k=1$ for sweeping values of $d_s$. The circles present the values given by Equation~\ref{eq:choosingDsIntermediate}}.
%\label{fig:multipleM} 
%\end{figure}

Let us again consider Term~$\mathcal{B}$ in Equation~\ref{eq:dMaxPureOrder1g} and replace $T_a$ using Equation~\ref{eq:choosingTa} and $\gamma$ by $d_s - d_a$.
Then, Term~$\mathcal{B}$  becomes
\begin{equation}
\label{eq:termBtmp}
\left\lceil\frac{T_s + (d_s - d_a)(1 - k_c)}{k_c \cdot (d_s - d_a)}\right\rceil \cdot \left\lceil \frac{k_c \cdot T_s}{T_s + d_s - d_a} \right\rceil.
\end{equation}
As long as the number of steps $\gamma$-steps (and hence the value of the first ceiling function in Equation~\ref{eq:termBtmp}) is not increased, we have the possibility to minimize $d_s$ or to maximize $T_s$. This will decrease the duty-cycle for reception, such that we can allocate more of the overall duty-cycle $\eta$ for transmission. This, in turn, will allow us to minimize $T_a$, to which $d_m$ is nearly proportional (cf. Equation~\ref{eq:dMaxPureOrder1g}). Large values of $T_s$ or small values of $d_s$ increase the term within the first ceiling function of Equation~\ref{eq:termBtmp}.
We therefore need to find the values of $T_s$ and $d_s$ that maximize this term, without causing the ceiling-function to turn over.
Hence, we require:
\begin{equation}
\label{eq:Ts_optimization_criterion}
\frac{T_s + (d_s - d_a)(1 - k_c)}{k_c(d_s - d_a)}  = M, M \in \mathbb{N}
\end{equation}
In other words, values of $T_s$ and $d_s$ that fulfill Equation~\ref{eq:Ts_optimization_criterion} lead to the same number of $\gamma$-steps (i.e. $M$ steps). As long as this equation is fulfilled, we can optimize the other terms of the latency equation without increasing the first ceiling function in Term~$\mathcal{B}$ in Equation~\ref{eq:dMaxPureOrder1g}.
Solving Equation~\ref{eq:Ts_optimization_criterion} by $T_s$ leads to the following relation between $T_s$ and $d_s$.
\begin{equation}
\label{eq:choosingTs}
T_s = (k_c (M + 1) - 1) (d_s - d_a)
\end{equation}
Inserting Equation \ref{eq:choosingTa} and \ref{eq:choosingTs} into Equation~\ref{eq:dMaxPureOrder1g} leads to
\begin{equation}
\label{eq:genericPIkMLatency}
d_{m}\!=\!
\begin{cases}
M(M+1)(d_s - d_a) + d_a,& \mbox{if } k_c = 1,\\
(M+1)(d_s - d_a) (k_c (M+1) - 1) + d_a, &\mbox{else.}
\end{cases}
\end{equation}
Further, Equation~\eqref{eq:etaDef} can be expanded to the following duty-cycle.
\begin{equation}
\label{eq:etaPIkMp}
\eta(d_s, k_c, M) = \frac{d_s - d_a + k_c d_a + M(d_s + k_c d_a)}{(d_s - d_a)(M + 1)(k_c + kM - 1)}
\end{equation}
We can rearrange this to
\begin{equation}
\label{eq:dsByEta_km}
d_s(\eta,k_c,M) = \frac{d_a \cdot (\eta + M \eta + 1)\cdot(k_c + M k_c - 1)}{(\eta (k_c + k_c M - 1) - 1)\cdot(M + 1)}
\end{equation}
Recall that $d_s$ is the length of the reception window, whereas $k_c$ and $M$ are two integer values that are yet to be determined.
We can replace $d_s$ in Equation~\ref{eq:dMaxPureOrder1g} by Equation~\ref{eq:dsByEta_km}. This gives us the worst-case latency depending on the duty-cycle. Compared to the initial situation, in which we had to choose 3 real-valued parameters $T_s$, $T_a$ and $d_s$, we have simplified the problem to optimizing two integer numbers $k_c$ and $M$.

\subsubsection{Optimizing ${k_c}$ and ${M}$}

We can easily find the value of $k_c$ that leads to a local latency minimum by solving ${\frac{d}{dk_c} \left(d_m(k_c,M,\eta)\right) = 0}$ using Equation \ref{eq:genericPIkMLatency} and \ref{eq:dsByEta_km}. The resulting optimal value of $k_c$ is given by:
\begin{equation}
\label{eq:PIkM-kOpt}
k_{opt} = \frac{1}{M+1} + \frac{\sqrt{\eta + M\eta + 1} + 1}{\eta(M+1)}
\end{equation}
Since $k_c$ needs to be an integer, we have to chose $k_c = \mbox{round}(k_{opt})$. For optimizing $M$, let us first consider duty-cycles $\eta$ for which $\mbox{round}(k_{opt}) = k_{opt}$ and compute the derivative ${\frac{d}{dM} (d_m(k_{opt},M,\eta)})$, which represents the slope of the worst-case latency. The resulting terms are positive for all possible $M$, and hence low values of $M$ lead to low latencies. 
Though this deviation is only possible for duty-cycles for which $\mbox{round}(k_{opt}) = k_{opt}$, we can generalize this towards all duty-cycles due to the following reason.

Recall that a packet being sent within the last $d_a$ time-units of every scan window cannot be received successfully (since it only partially coincides with the scan window). Nevertheless, also the last $d_a$ time-units of every scan window contribute to the duty-cycle. We can write the worst-case latency of the MultiInt-Case from Equation~\ref{eq:genericPIkMLatency} for $k_c > 1$ as $d_m = (M+1) \cdot T_s + d_a$. It follows that $M$ determines within how many scan intervals the discovery will be successful. 
For higher values of $M$, the duty-cycle for reception is spread over an increasing number of scan windows per $d_m$. Hence, also the share of ``unproductive'' versus ``productive'' reception time is increased, leading to a higher latency for a given duty-cycle.

As a result, $M=1$ is the value with the highest performance, whereas all larger values of $M$ will increase worst-case latency. We will therefore first analyze the case of $M = 1$ in detail. As we will describe in Section~\ref{sec:implementation}, the case of $M=2$ has beneficial properties to handle non-idealities of the radio in symmetric scenarios, with only a marginally increased latency compared to $M = 1$. We therefore also study the case of $M=2$ in detail.

\subsubsection{Constraints for ${M = 1}$}
In the following, we assume $M=1$ and analyze the resulting properties. 
Though we can determine the theoretically optimal value of $k_c$ using Equation~\ref{eq:PIkM-kOpt}, not all of these optimal values are feasible. First, $d_s(\eta)$ needs to be positive.
This is equivalent to requiring a positive denominator of $d_s$ given by Equation~\ref{eq:dsByEta_km}, which leads to the following constraint.
\begin{equation}
\label{eq:PIk1_constraintA1}
k_c > \frac{1}{2 \eta} + \frac{1}{2}.
\end{equation}
In addition, since real-world hardware cannot realize a scan window that is shorter than a certain threshold $d_{s,m}$, we require $d_s(\eta) \ge d_{s,m}$.
By expressing $d_s$ using Equation~\ref{eq:dsByEta_km} and by solving this inequality, one can derive that $k_c$ needs to fulfill the following constraints.
\begin{equation}
\label{eq:PIk1_constraintB}
\begin{array}{lcl}
k_c \le k_l, &\mbox{if}& \eta > \frac{d_a}{2(d_{s,m} - d_a)},\\
k_c \ge k_l, &\mbox{if}& \eta < \frac{d_a}{2(d_{s,m} - d_a)},\\
\end{array}
\end{equation}
with \begin{equation}
\label{eq:PIk1_kl}
k_{l} = \frac{2 d_{s,m} (1+\eta) - d_a(1+2\eta)}{4\eta d_{s,m} - 2 d_a (1 + 2\eta)}.
\end{equation}
For $\eta = \frac{d_a}{2(d_{s,m} - d_a)}$, no constraints on $k_c$ apply.

Therefore, $k_c$ is chosen as $\mbox{round}(k_{opt})$ (cf. Equation~\ref{eq:PIkM-kOpt}), or as the value being closest to $\mbox{round}(k_{opt})$ allowed by these constrains. When comparing Equations \ref{eq:PIk1_constraintA1} and \ref{eq:PIk1_constraintB}, one can infer that the maximum duty-cycle that can be realized when $M=1$ is given by
\begin{equation}
\label{eq:PIk1_etaLimit}
\eta \le \frac{3 d_a + \sqrt{d_a (d_a + 8 d_{s,m})}}{8 (d_{s,m} - d_a)}.
\end{equation}
Next, we examine parametrizations with $M = 2$.

\subsubsection{Constraints for $M = 2$}
As for $M=1$, $k_c$ needs to be set to $\mbox{round}(k_{opt})$, where $k_{opt}$ is given by Equation~\ref{eq:PIkM-kOpt}. However, there are multiple constraints on $k_c$. First, $d_s(\eta)$ needs to be positive, and from Equation~\ref{eq:dsByEta_km} follows that
$k_c > \nicefrac{1}{(3 \eta)} \cdot (\eta + 1) = k_{min}.$
Second, there is a lower limit $d_{s,m}$ on the scan window the hardware can realize, as already described. One can solve $d_s(\eta) > d_{s,m}$ by $k_c$ using Equation \eqref{eq:etaPIkMp}, which results in the following upper limit on $k_c$.
\begin{equation}
\label{Eq:kmax}
k_c \leq \frac{d_{s,m}}{3 \eta d_{s,m} - (3 \eta + 1) d_a} + \frac{1}{3} = k_{max}
\end{equation}
Since k is an integer value, $\lceil(k_{min}) \rceil \leq \lfloor k_{max} \rfloor$ must always be kept. A conservative but analytically solvable form of this inequality is $k_{min} + 1 \leq k_{max} - 1$.
By solving this using $k_{min}$ and $k_{max}$ as given above, we get an upper limit on the duty-cycle that can be realized:
\begin{equation}
\label{eq:etamaxpikmp}
\eta \leq \frac{3 d_a + \sqrt{d_a (d_a + 8 d_{s,m})}}{12(d_{s,m} - d_a)}
\end{equation}

\subsection{Other Parametrizations}
\label{sec:otherParametrizations}
The presented parametrizations had in common that $0 < \gamma \leq d_s - d_a$. We in this section discuss the remaining parametrizations and why they do not perform better than the ones already described.
First, let us consider parameters with $\gamma = 0$, e.g., $T_a = 2 \cdot T_s$. Here, the offset $\Phi_i$, i = 1,2,... (or $\Phi_i^\prime$, respectively) between a beacon and its neighboring scan window will always remain constant. Hence, the temporal distance between beacons and windows is not decreased over time, and for certain ranges of initial offsets, the devices never discover each other. In other words, parametrizations with $\gamma = 0$ are not suited for guaranteeing discovery within bounded time.
\begin{figure}[tb]
	\centering
	\includegraphics[width=\linewidth]{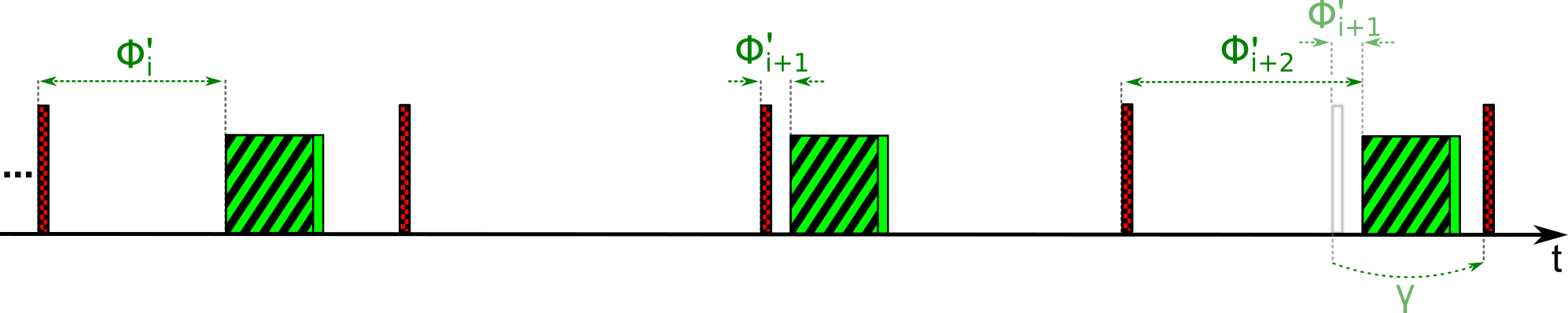}
	\caption{Growing situation with $\gamma > d_s - d_a$. Here, $\Phi_{i+2}^\prime > \Phi_{i+1}^\prime$.}
	\label{fig:gammagtdsmda} 
\end{figure}

We next discuss parametrizations using which $\gamma$ exceeds $d_s - d_a$ time-units. Figure~\ref{fig:gammagtdsmda} depicts a growing sequence with $\gamma > d_s - d_a$. The blurred beacon in front of the last scan window has an offset of $\Phi_{i+1}^\prime$. This offset is equal to that of the beacon preceding the second scan window. As can be seen, since $\gamma$ exceeds $d_s - d_a$, this beacon is ``shifted'' on the right side of the scan window for certain offsets, such that it is $\Phi_{i+2}^\prime > \Phi_{i+1}^\prime$. Therefore, parametrizations with $\gamma > d_s - d_a$ cannot guarantee discovery by reducing the offset in multiple steps of length $\gamma$. Nevertheless, when extending such a sequence by considering future beacons and scan windows, discovery might be guaranteed for all offsets at a later point in time. This works follows.

For the \textit{MultiInt} scheme in case b), we have considered one instance of $T_s$ and $k_f = \left\lfloor\nicefrac{T_s}{T_a}\right\rfloor$ instances of $T_a$ for computing  $\gamma = T_s - k_f \cdot T_a$. We now consider a multiple $l$ of $T_s$, such that $\gamma^\prime = l \cdot T_s - k_f^\prime \cdot T_a$. If we can identify a tuple $(l, k_f^\prime)$ for which $\gamma^\prime \leq d_s - d_a$, then the distance between a scan window and its appropriate neighboring beacon shrinks by $\gamma^\prime$ time-units after every $k_f^\prime$ advertising intervals, and a beacon always overlaps with a scan window after a finite number of such steps.
It has been shown in \cite{kindt:15} that suitable values of $(l, k_f^\prime)$ exist for almost every value of $T_a$ and $T_s$.

Hence, it is possible to construct parametrizations schemes with $\gamma > d_s - d_a$, if $\gamma^\prime$, which is formed by multiples of $T_a$ and $T_s$, does not exceed $d_s - d_a$.
The properties of such sequences are similar to the ones with $\gamma \leq d_s - d_a$. In particular, a finite number of steps of length $\gamma^\prime$ are needed until discovery is guaranteed. However, more scan windows need to take place for reaching the same worst-case latency as parametrizations with $\gamma \leq d_s - d_a$ (which can be achieved by reducing $T_s$). 
This leads to a higher fraction of ``unproductive'' scanning time within the last $d_a$ time-units of each scan window, and hence the duty-cycle for guaranteeing the same worst-case latency is increased.  In conclusion, it is not possible to further increase the performance with parametrizations with $\gamma > d_s - d_a$. 
Furthermore, parametrizations in which no $\gamma^\prime \leq d_s - d_a$ exists do not provide bounded latencies.

%% file: sections/proof_of_optimailty.tex
\subsection{Proof of Optimality}
\label{sec:proofOfOptimality}
In this section, we formally prove that the discovery latencies achieved by PI-based protocols parametrized using the SingleInt scheme are actually optimal.
In particular, when considering one sender and one receiver, for a given sum of duty-cycles (and hence energy budgets) of both devices, no other \ac{ND} protocol can guarantee lower worst-case latencies than a PI-based one configured according to the SingleInt scheme. We then discuss symmetric bi-directional discovery and the MultiInt scheme.

In  this section, for the sake of simplicity of exposition only, we neglect the transmission duration of the successfully received beacon, which is small compared to $d_m$. In addition, we assume that also packets that partially overlap with the last $d_a$ time-units of a reception window are received successfully. Our proof of optimality also holds true when relaxing these assumptions (cf. ~\cite{kindt:19} for details).
\subsubsection{Performance Bound for Unidirectional Discovery}
\begin{figure}[t!]
	\centering
	\includegraphics[width=\linewidth]{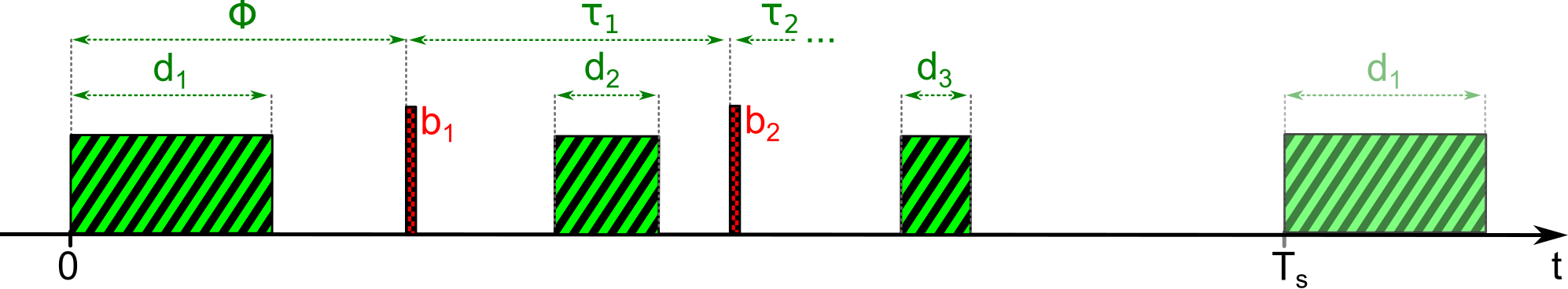}
	\caption{Sequence of reception windows of length $d_1$, $d_2$, $d_3$, into which the sequence of beacons $b_1$, $b_2$ falls with a random offset $\Phi$.}
	\label{fig:optimality_situation} 
\end{figure}

We now derive the lowest worst-case latency that any \ac{ND} protocol could guarantee between one sender with a duty-cycle of $\beta$ and one receiver with a duty-cycle of $\rho$. Additional details of this derivation, and a more formal proof, can be found in~\cite{kindt:19}.

The reception duty-cycle $\rho$ is the fraction of time during which the radio listens to the channel. A \ac{ND} protocol could use an arbitrary pattern of reception windows. We assume that this pattern repeats after a certain (arbitrary) period $T_s$. The sum of reception window lengths during every period $T_s$ is given by $\sum_i d_i$, and hence, $\rho = \nicefrac{(\sum_i d_i)}{T_s}$. Such a pattern is exemplified in Figure~\ref{fig:optimality_situation}. Here, three windows of length $d_1$, $d_2$ and $d_3$ are repeated with a period of $T_s$. These windows could be spaced arbitrarily. 
Let us for now assume that a given pattern is already optimal. Our goal is to identify a corresponding optimal sequence of beacons, such that one of these beacons coincides with a reception window within the shortest possible time in the worst-case.

When the sender transmits its first beacon $b_1$ after both devices have come into range, as depicted in Figure~\ref{fig:optimality_situation}, it will fall with a random offset $\Phi \in [0, T_s]$ into an instance of the pattern of reception windows (cf. Figure~\ref{fig:optimality_situation}). 
With what probability $P_r$ does $b_1$ overlap with a reception window? Clearly, $P_r = \nicefrac{(\sum_i d_i)}{T_s} = \rho$.
Next, the sender transmits its next beacon $b_2$ after $\tau_1$ time-units. What is the probability that either $b_1$ or $b_2$ is received? 
Obviously, the probability depends on the value of $\tau_1$ and the pattern of reception windows. However, if this pattern and the value of $\tau_1$ are chosen in the best possible way, the offsets for which $b_2$ coincides with a reception window are disjoint with those of $b_1$. Hence, the probability that either of them coincides with a reception window is given by $P_r^\prime = \nicefrac{2 \cdot (\sum_i d_i)}{T_s} = 2 \cdot \rho$.
Note that otherwise, if for some offsets $\Phi$, both $b_1$ and $b_2$ would \textit{redundantly} overlap with a reception window, the fraction of successful offsets and hence $P_r^\prime$ would be reduced.

If we now continue the sequence $b_1, b_2, b_3,...$ by sending additional beacons, such that for no offset, more than one beacon overlaps with a reception window, how many beacons $N$ need to be sent until the probability that at least one of them overlaps becomes unity? Clearly, we need to fulfill $N \cdot \rho \geq 1$. From this follows that any \ac{ND} protocol needs to send at least $N$ beacons until discovery can be guaranteed for all offsets $\Phi$, and $N$ is given by $N = \left \lceil \nicefrac{1}{\rho}\right\rceil$.

How should such $N$ consecutive beacons be spaced, i.e., what values of $\tau_1$, $\tau_2$,... are optimal? 
A duty-cycle for transmission of $\beta$ implies that on the average, every two consecutive beacons must be spaced by $\lambda = \nicefrac{d_a}{\beta}$ time-units. 
More precisely, when any $N$ consecutive beacons are spaced by less than $(N-1)\cdot \lambda$ time-units, a later sequence of $N$ consecutive beacons needs to be spaced by more than $(N-1)\cdot \lambda$ time-units for maintaining the duty-cycle $\beta$. Every beacon could be the first beacon sent after two devices have come into range, and the longest sequence of $N$ consecutive beacons determines the worst-case discovery latency. As a result, in an optimal \ac{ND} protocol, every $N$ consecutive beacons are spaced by $(N-1) \cdot \lambda$ time-units.
The worst-case beacon-to-beacon discovery latency is hence given by $d_m^* = (N-1) \cdot \lambda$ time-units, and the worst-case discovery latency after two devices have come into range is given by $N \cdot \lambda$ time-units. Expressing $N$ and $\lambda$ by $\beta$ and $\rho$ leads to the lowest worst-case discovery latency that any \ac{ND} protocol could guarantee~\cite{kindt:19}.
\begin{equation}
\label{eq:BoundUnidir}
d_m = \left\lceil \frac{1}{\rho}\right\rceil \cdot \frac{d_a}{\beta}
\end{equation}
Since $\eta = \beta + \rho$, we have the flexibility to adjust $\beta$ and $\rho$ for a given $\eta$, i.e., to optimize the allocation of the energy budget between the sender and the receiver.
When substituting $\beta = \eta - \rho$ in Equation~\ref{eq:BoundUnidir}, one can verify that only values of $\rho$ for which $\nicefrac{1}{\rho} = k,\ k = 1,2,3,...$ can minimize $d_m$. 
By analyzing the slope of $d_m$, we can show that $k = \nicefrac{2}{\eta}$ optimizes $d_m$, which is in general a non-integer value and hence cannot be realized. From the monotonicity of $d_m$, it follows that only the neighboring pair of integers $k_l = \lfloor \nicefrac{2}{\eta} \rfloor$ and $k_r = \lceil \nicefrac{2}{\eta} \rceil$ can minimize $d_m$ (cf.~\cite{kindt:19} for details). Substituting $\nicefrac{1}{\rho}$ by $k_l$ and $k_r$ in Equation~\ref{eq:BoundUnidir} results in the following bound.
\begin{equation}
\label{eq:symBound}
d_m = \min\Bigg(\left\lceil\frac{2}{\eta}\right\rceil^2 \cdot \frac{d_a}{\eta \left\lceil\frac{2}{\eta}\right\rceil - 1},\mbox{ }\left\lfloor\frac{2}{\eta}\right\rfloor^2 \cdot \frac{d_a}{\eta \left\lfloor\frac{2}{\eta}\right\rfloor - 1}\Bigg)
\end{equation}
No \ac{ND} protocol can guarantee a lower discovery latency between a pair of a sender and receiver for a given joint duty-cycle $\eta$ than that given by Equation~\ref{eq:symBound}.
We next show that the SingleInt scheme realizes the discovery latencies given by Equation~\ref{eq:symBound}.

\subsubsection{Optimality of the SingleInt Scheme}
When assuming i)~that beacons are successfully received if they are sent within the last $d_a$ time-units of a reception window and ii) neglecting the transmission duration of the successfully received beacon, we can set $d_a = 0$ in Equation~\ref{eq:dMaxPureOrder0} and obtain the following worst-case latency for the SingleInt scheme:
\begin{equation}
\label{eq:equationPIkmRaw}
d_m = \left ( \left \lceil \frac{T_s - d_s}{T_a} \right \rceil + 1\right)\cdot T_a
\end{equation}

Furthermore, Equations~\ref{eq:0M-Ta} and \ref{eq:choosingTs_expanded} become  $T_a = d_s$ and ${T_s = (M+1) d_s}$, $M \in \mathbb{R}$ under these assumptions. When inserting this into Equation~\ref{eq:equationPIkmRaw}, we obtain $d_m = (M+1)\cdot d_s$. We can replace $d_s$ using Equation~\ref{eq:etaDef} and obtain a worst-case latency $d_m$ of $\frac{d_a (M + 1)^2}{\eta (M+1)-1}$ time-units. Note that despite the assumptions described above, $d_a$ must not be set to zero in Equation~\ref{eq:etaDef}, since beacon transmissions still contribute to the duty-cycle.
The derivative $\nicefrac{d}{d_M}\:d_m$ is as follows.

\begin{equation}
\label{eq:dmdMproof}
\frac{d\:d_m}{d\:M} = \frac{d_a (M+1) \cdot (\eta + M\cdot \eta - 2)}{(\eta + M\cdot \eta - 1)^2}
\end{equation}
The optimal value of $M$ is obtained by computing the local minimum of $d_m$, which can be done by solving $\nicefrac{d}{d_M}\:d_m = 0$ by $M$. 
By setting the numerator in Equation~\ref{eq:dmdMproof} to zero, we obtain an optimal value of $M_o = \nicefrac{2}{\eta} - 1$. We can verify that this value indeed corresponds to a minimum by observing the second derivative. Since $M$ must be an integer value (cf. Section~\ref{sec:PI_0M}), we identify the neighboring integers $M_l = \lfloor M_{o} \rfloor$ and 
$M_r = \lceil M_{o} \rceil$. When rounding $M_o$ up to its next higher (i.e., $M = M_r$) or lower (i.e., $M = M_l$) integer-value, the following latency is obtained.
\begin{equation}
\begin{array}{lr}
d_{m,r} = \left\lceil \frac{2}{\eta}\right\rceil^2 \cdot \frac{d_a}{\eta \left\lceil\frac{2}{\eta}\right\rceil - 1},&
d_{m,l} = \left\lfloor \frac{2}{\eta}\right\rfloor^2 \cdot \frac{d_a}{\eta \left\lfloor\frac{2}{\eta}\right\rfloor - 1}
\end{array}
\end{equation}
If we now set $M = \lceil M_l \rceil$, if $d_{m,l} \leq d_{m,r}$, or  $M = \lfloor M_r \rfloor$ otherwise, the resulting latency will always be identical to $\min(d_{m,l},d_{m,r})$. This worst-case latency is identical to the latency bound given by Equation~\ref{eq:symBound}. Hence, the SingleInt scheme (with the described modification of setting $M$ to $M_l$ or $M_r$ instead of $round(M_o)$) performs optimally. 
We next discuss symmetric, bi-directional discovery.
\subsubsection{Performance Bound for Symmetric Bidirectional Discovery}
Let us now  assume that each device both transmits with a duty-cycle of $\beta$ and receives with a duty-cycle of $\rho$. Hence, every device now has a duty-cycle of $\eta = \beta + \rho$, and $\eta$ is partitioned into $\beta$ and $\rho$ as described for the unidirectional case. As long as no beacons collide, the worst-case discovery latency of both devices discovering each other is given by Equation~\ref{eq:symBound}. Hence, the latency bound reached in \textit{most} cases is equal to that of the unidirectional case, but due to colliding beacons, a certain fraction of attempts exceed this bound. As we will describe in Section~\ref{sec:implementation}, because of this and other effects, the rate of failed discoveries is non-negligible for the bi-directional SingleInt case. We however propose a variant of the MultiInt scheme in Section~\ref{sec:implementation}, which provides latencies only marginally above that of the SingleInt scheme, while discovery is successful in $\SI{99.997}{\percent}$ to $\SI{99.804}{\percent}$ of all attempts when a pair of devices discover each other bi-directionally.
\subsubsection{Summary}
In summary, for one sender and receiver, the SingleInt scheme is optimal, and no other \ac{ND} protocol can guarantee lower discovery latencies. The MultiInt scheme provides latencies slightly above that of the SingleInt scheme and is hence only near-optimal.
For symmetric bi-directional discovery, no \ac{ND} protocol can guarantee bounded latencies in $\SI{100}{\percent}$ of all attempts~\cite{kindt:19}. As of today, which parametrization or \ac{ND} protocol optimizes the trade-off between discovery latency, failure rate and energy consumption for a given number of devices is not known.

%% file: sections/implementation.tex
\section{Implementation}
\label{sec:implementation}	
This section describes the implementation of PI-based protocols that use the presented parametrization schemes.
We first consider one-way discovery, i.e., scenarios with one transmitter and one receiver. As will become clear later, all of our proposed protocol variants are suitable for one-way discovery, but the SingleInt scheme provides the lowest latencies. We then study symmetric two-way discovery, where a pair of devices discover each other simultaneously. Only the MultiInt scheme suits such scenarios, since the SingleInt scheme leads to a large fraction of failures. We therefore present a practical implementation of the MultiInt scheme for symmetric two-way discovery, which accounts for non-idealities of the hardware. Finally, we study how our proposed schemes can be used to parametrize BLE.
\subsection{Range of Duty-Cycles}
\begin{figure}[tb]
	\centering
	\includegraphics[width=\linewidth]{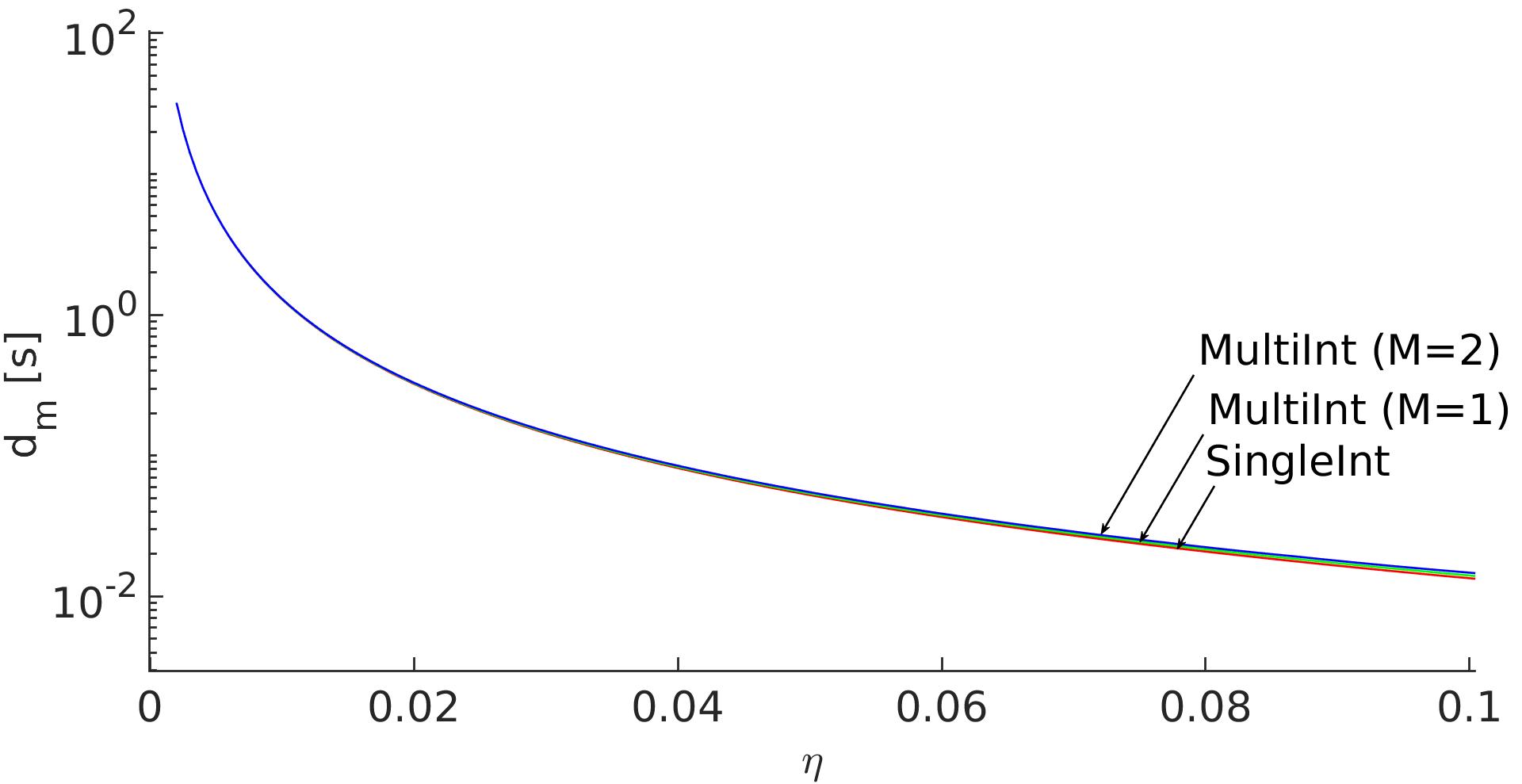}
	\caption{Worst-case latencies of our proposed schemes for different duty-cycles ($d_a = 32 \mu s$).}
	\label{fig:wcLatencies} 
\end{figure}

Figure \ref{fig:wcLatencies} depicts the computed worst-case latencies of the three proposed parametrization schemes (viz., SingleInt (\textit{SI}), MultiInt (\textit{MI}) with M=1 and M=2), obtained from the equations presented in the previous section. 
We assume a packet transmission duration $d_a$ of $\SI{32}{\micro s}$, which we will justify in Section~\ref{sec:evaluation}.
As can be seen in Figure~\ref{fig:wcLatencies}, especially for low duty-cycles, all of our proposed parametrizations perform almost identically, with the SingleInt scheme providing slightly lower worst case latencies for larger duty-cycles. This difference in performance is caused by the different shares of ``unproductive'' duty-cycle, as explained in the previous section.

In the rest of this section, we restrict our considerations to duty-cycles between $\SI{0.2}{\percent}$ and $\SI{1.55}{\percent}$, since they lead to a range of worst-case latencies that is relevant in practical applications.
In particular, $\eta = \SI{0.2}{\percent}$ corresponds to a worst-case latency of roughly half a minute, $\eta = \SI{1.55}{\percent}$ to roughly half a second. Especially for this range of duty-cycles, all of the three parametrization schemes perform nearly identically (cf. Figure~\ref{fig:wcLatencies}).
%\begin{figure}[htb]
%	\centering
%	\includegraphics[width=\linewidth]{images/intervalLengths_genuine.png}
%	\caption{Parameter values for $T_a$, $T_s$ and $d_s$ for different duty-cycles. \textit{SI} represents the SingleInt scheme, %\textit{MI} for the MultiInt Scheme.}
%	\label{fig:intervalLengths} 
%\end{figure}
\begin{table}[tb]
\small
\centering
{\setlength{\tabcolsep}{0.185cm}
\renewcommand{\arraystretch}{1.1}
\begin{tabular}{|c|ccc|ccc|}
\hline
&\multicolumn{3}{c|}{SingleInt}&\multicolumn{3}{c|}{MutliInt\ (M=2)}\\
\hline
$\eta [\%]$ & $T_a [s]$ & $T_s [s]$ & $d_s[s] $ & $T_a [s]$ & $T_s[s]$ & $d_s[s]$\\
\hline
0.20 & 0.0320 & 32.0320 & 0.0321 & 0.0321 & 10.6986 & 0.0107 \\
0.55 & 0.0117 & 4.2430 & 0.0117 & 0.0117 & 1.4221 & 0.0039 \\
0.90 & 0.0071 & 1.5874 & 0.0072 & 0.0071 & 0.5338 & 0.0024 \\
1.20 & 0.0054 & 0.8942 & 0.0054 & 0.0054 & 0.3016 & 0.0018 \\
1.55 & 0.0041 & 0.5369 & 0.0042 & 0.0042 & 0.1817 & 0.0014 \\
\hline
\end{tabular}
}
\caption{Parameter values for $T_a$, $T_s$ and $d_s$ for different duty-cycles.}
\label{tab:intervalLengths}
\end{table}
Table~\ref{tab:intervalLengths} shows the interval lengths computed by the equations presented in the previous section. The advertising intervals are essentially identical for both schemes and lie between $\SI{4.1}{ms}$ and $\SI{32.1}{ms}$ for the considered range of duty-cycles. The smallest scan window length lies around $\SI{1.4}{ms}$. The scan intervals are considerably larger, reaching up to $\SI{32}{s}$ for the SingleInt scheme. 
This range of values can be realized easily by available hardware. Note that these interval lengths become larger, if more bytes per beacon are sent, as required by the \ac{BLE} standard. We study \ac{BLE} in Section~\ref{sec:ble}, present the resulting ranges of values and conclude that they comply with the \ac{BLE} standard.

Even though PI-based protocols can be implemented easily by using hardware timers for scheduling $T_a$ and $T_s$, the hardware properties of the radios, such as turnaround times or clock inaccuracies, impose challenges on their implementation. In the following, we analyze these challenges in detail and propose countermeasures to overcome them.

\subsection{One-Way Discovery}
Implementing one-way discovery using our proposed parametrizations is relatively simple. However, we have to take two measures to account for non-idealities of the hardware, i.e., incorporating a safety margin into $T_s$ and compensating for clock quantization errors.
\subsubsection{Safety Margin of $T_s$}
In Equations~\ref{eq:choosingTsRelation} and~\ref{eq:Ts_optimization_criterion}, we have chosen $T_s$ such that the latency function will increase abruptly for every smallest increase of $T_s$ (since $T_s$ lies directly in front of a step of the $d_m$-function). On real-world hardware, we have to ensure that $T_s$ never exceeds its computed optimal value, despite of clock skew and other errors. Therefore, $T_s$ needs to be set slightly smaller than its computed optimal value, i.e., by one tick of the radio's sleep oscillator.
\subsubsection{Clock Quantization Error Correction}
Our proposed parametrizations will lead to a large number of (short) advertising intervals until discovery is guaranteed. For example, for $\eta = \SI{0.2}{\percent}$, in the worst-case,  around $1000$ instances of $T_a$ will pass until discovery occurs in  each of the proposed schemes. The sleep oscillators of most wireless devices run on very low frequencies for maintaining energy-efficiency. For example, the clock frequency $f_{clk}$ is $\SI{32768}{Hz}$ for the nRF51822-radio considered in this paper~\cite{nrf51822:14}. This limits the granularity of $T_a$ and $T_s$ to $1 \cdot T_{clk} \approx \SI{30.5}{\micro s}$, and quantization errors can pile up to $\SI{30.5}{\milli s}$ after $1000$ advertising intervals in the worst-case. As a result, the effective value of the offset-shrinkage or growth per scan interval, $\gamma$, deviates from its optimal value of $d_s - d_a$ and hence, the deterministic overlap of beacon sequences with their corresponding scan windows is not guaranteed anymore. 
For e.g., $\eta = \SI{0.2}{\percent}$ in the SingleInt case, $d_s - d_a = \SI{32.1}{\milli s}$ and hence, the maximum accumulated quantization error of $\SI{30.5}{\milli s}$ is comparable to this. 
As our experiments confirmed, this leads to a high number of cases in which the predicted worst-case latencies are significantly exceeded.
To overcome this problem, each device must remember the exact values of its advertising- and scan interval in its memory, with a high precision of e.g., $\SI{1}{\nano s}$. Whenever the sleep clock wakes up the CPU, the accumulated quantization error $Q$ is computed by calculating the difference of the time that had passed according to the number of clock ticks and the time that should have passed, based on the exact interval lengths stored in memory. 
As soon as $|Q|$ exceeds $1/2 \cdot T_{clk}$, the next interval instance is extended or shortened by 1 clock tick, which is again taken into account for computing the next value of $Q$. 
With this technique, the resulting effective value of $\gamma$ cannot deviate by more than ${1 \cdot T_{clk}}$ from its optimal value due to quantization errors. To compensate for the remaining error, $d_s$ needs to be extended by at least $1 \cdot T_{clk}$ beyond its ideal value. We assume an extension of $d_s$ by $5 \cdot T_{clk}$ to account for remaining inaccuracies (i.e., quantization errors and clock skew) in our Evaluation (cf. Section~\ref{sec:evaluation}).

\subsection{Symmetric Two-Way Discovery}
\label{sec:symmetric_2w}
We now extend the one-way discovery scenario to symmetric two-way discovery. In principle, each device schedules both beacons and reception windows using the parameter values described in Section~\ref{sec:PIkMProtocol}. We assume that both devices use the same duty-cycle $\eta$ (i.e. symmetric neighbor discovery).

\begin{figure}[b]
	\centering
	\includegraphics[width=\linewidth]{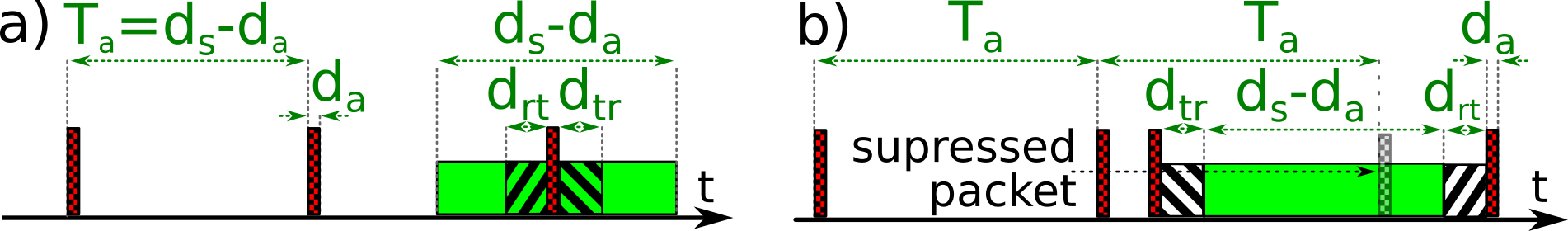}
	\caption{a) Blockage of the SingleInt scheme and b) blockage-compensated MultiInt scheme with $M=2$. In the hatched areas, the radio is unable to receive or transmit. These areas lie within the reception window in a) and outside in b).}
	\label{fig:txrxblocking} 
\end{figure}

In real-world implementations, beacon collisions and non-negligible durations for switching from reception to transmission and vice-versa are always present, as shown in Figure \ref{fig:txrxblocking}. The rectangle in Figure~\ref{fig:txrxblocking}a) depicts a scan window and the hatched bars multiple beacons of the same device, which are scheduled according to the SingleInt scheme.

In the SingleInt scheme, we apply configurations with $T_a \leq d_s - d_a$ to guarantee that a beacon overlaps with \textit{every} scan window of a remote device. This, however, also implies that a beacon will overlap with every reception window of the same device (cf. Figure~\ref{fig:txrxblocking}a)).
In addition, every radio requires a certain amount of time for switching from reception to transmission ($d_{rt}$) and vice-versa ($d_{tr}$). For the radio we consider, each of these durations spans approximately $\SI{140}{\micro s}$ \cite{nrf51822:14}, which makes them large compared to the packet transmission duration of $\SI{32}{\micro s}$. Hence, within $d_{rt}$, $d_a$ and $d_{tr}$ time-units, the radio is unable to receive any incoming packets. One can verify from Figure~\ref{fig:txrxblocking} that no matter by which amount of time the sequence of transmissions is shifted against the sequence of reception windows on the same device, $d_{rt} + d_a + d_{tr}$ time-units are blocked within every scan window.

In other words, there are some offsets $\Phi_0$ between the beacon and reception patterns of two devices, for which no beacon is received successfully.
The mean blockage probability of the SingleInt-scheme among all possible offsets $\Phi_0$ is
\begin{equation}
\label{eq:PI0Mblockage}
p_{blk} = \frac{d_{rt} + d_a + d_{tr}}{d_s - d_a}.
\end{equation}
Within our considered range of duty-cycles between $\SI{0.2}{\percent}$ and $\SI{1.55}{\percent}$, the smallest scan window length according to the SingleInt scheme is $\SI{4.2}{\milli s}$, which corresponds to an unacceptably high blocking probability $p_{blk}$ of $\SI{7.5}{\percent}$.
Unfortunately, there is no feasible way of mitigating this for the SingleInt parametrization scheme. However, we can effectively mitigate blocking in the MultiInt scheme, as we describe next.

\subsection{Symmetric Discovery using the MultiInt Scheme}
We can write Equation~\ref{eq:genericPIkMLatency} for $k_c>1$ as ${d_m = (M+1) \cdot T_s + d_a}$. Hence, a larger value of $M$ increases the number of scan intervals 
within which discovery is guaranteed. A beacon will only overlap with every (M+1)'th scan window and hence, blocking occurs only for a fraction of them. Recall that the discovery latency also increases with $M$, and hence a trade-off between low blocking probabilities and low discovery latencies need to be be achieved. We in the following consider $M = 2$, since it almost provides the same worst-case latencies as the SingleInt scheme, but can achieve extremely low blocking probabilities using the measures described next. 

\subsubsection{Blocking Mitigation}
\label{eq:blockMitiation}
To reduce the probability of blockage, we propose to suppress the transmission of every beacon that lies within a scan window.
To compensate for the omitted beacons, we send an additional beacon by $d_{tr}$ time-units before and another additional one by $d_{rt}$ time-units after every scan window.
The resulting beacon flow is shown in Figure~\ref{fig:txrxblocking}~b). As can be seen, no transmissions take place throughout the entire scan window. The additional beacons compensate for the omitted ones for the vast majority of offsets. One can derive that the remaining probability of a failed discovery is
\begin{equation}
\label{eq:PIk2blockage}
p_{blk} =  \frac{1}{2} \cdot \left(\frac{(d_{tr} + d_a)^2}{T_a T_s} + \frac{(d_{rt} + d_a)^2}{T_a T_s}\right)   +  \frac{d_{rt}+ d_{tr} + 2 d_a}{T_s}.
\end{equation}
This equation accounts both for the turnaround times and beacon collisions between two devices. Within the considered range of duty-cycles, the blocking probability is $\SI{0.003}{\percent}$ for $\eta = \SI{0.2}{\percent}$ and $\SI{0.193}{\percent}$ for $\eta  = \SI{1.55}{\percent}$. Figure~\ref{fig:piKM_blocking} depicts the probability of failed discoveries for all duty-cycles under consideration. However, due to the increased duty-cycle for sending two additional beacons, the worst-case latency for a given target duty-cycle $\eta$ is increased relatively by $\SI{0.6}{\percent}$ for $\eta=\SI{0.2}{\percent}$ and by $\SI{4.4}{\percent}$ for $\eta = \SI{1.55}{\percent}$. Due to the higher number of compensation beacons needed for $M=1$ and for the SingleInt-scheme, the MultiInt-scheme with $M=2$ is more beneficial for applying this technique.

We refer to the blocking-compensated version of the MultiInt scheme as \textit{MultiInt-BC}. Its differences  in a symmetric setting, compared to the uncompensated MultiInt scheme in a one-way scenario, are as follows.
\begin{compactitem}
	\item The worst-case latency is increased by up to $\SI{4.4}{\percent}$. 
	\item This worst-case latency can only be guaranteed in a large fraction of all discovery attempts (i.e., more than $\SI{99.8}{\percent}$), while $d_m$ might be exceeded in the remaining attempts.
\end{compactitem}
Note that the phenomenon of blocking occurs in all symmetric \ac{ND} protocols~\cite{kindt:19}. We will compare the worst-case latencies achieved by multiple \ac{ND} protocols with the same failure 	probability in Section~\ref{sec:evaluation}.
We next study parametrizations for the BLE protocol.

%% file: sections/ble.tex
\subsection{Configuring Bluetooth Low Energy}
\label{sec:ble}
Our proposed parametrization schemes can be tuned for parametrizing the BLE protocol, such that the discovery latency for a given joint duty-cycle, which is the sum of duty-cycles of both devices, is minimized.
In particular, the SingleInt scheme can be adopted to be fully compliant to the BLE specification, whereas the MultiInt scheme requires changing the range of random delays, which does not comply to the BLE standard.
Since the specification of BLE does not propose any optimized parameter values, to the best of our knowledge, we propose the first known efficient, closed-form parametrization schemes for BLE. 
Because BLE requires larger packet lengths, also the corresponding interval lengths become larger, making clock quantization error correction less relevant.
We provide a performance analysis of BLE configured using the SingleInt scheme in Section~\ref{sec:ble_performance}. 

\subsection{Adapting the SingleInt Scheme to BLE}
Recall that the SingleInt scheme is prone to failed discovery attempts in symmetric two-way scenarios. However, the problem of blocking does not occur in one-way scenarios, in which one device broadcasts packets without receiving, whereas the other device only receives without transmitting. Such scenarios are defined by the \textit{non-connectable undirected} advertising procedure of BLE \cite{bleSpec50}. We in the following first describe how the SingleInt scheme can be applied to BLE in such one-way scenarios, and then extend it to two-way scenarios.

\subsubsection{Non-Connected Undirected Advertising}
\begin{figure}[t!]
\centering
\includegraphics[width=\linewidth]{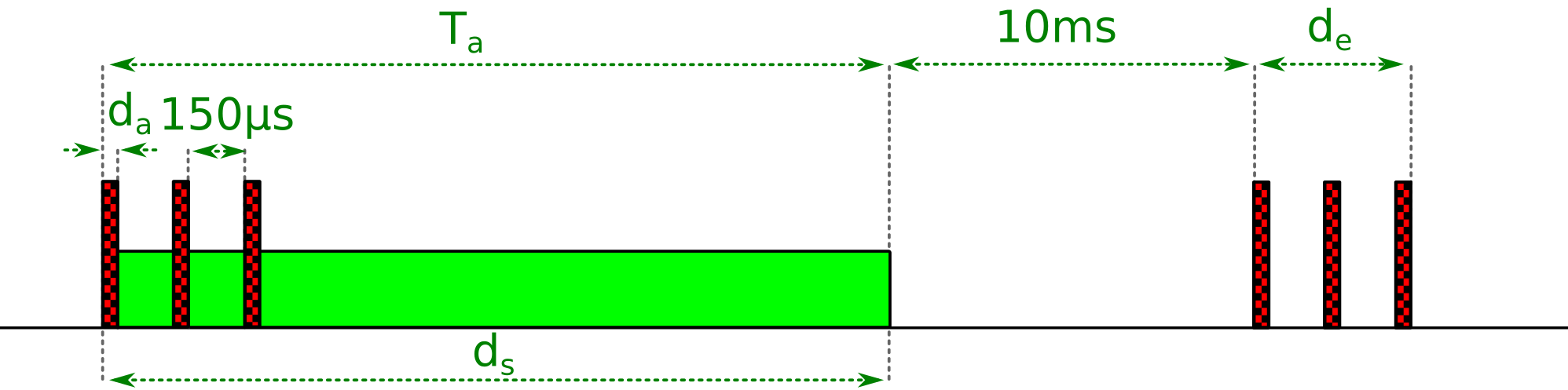}
\caption{The effective advertising interval in BLE is extended by up to $\SI{10}{ms} + d_e$ time-units.}
\label{fig:ble} 
\end{figure}

As already mentioned, BLE adds a random delay between $0$ and $\SI{10}{ms}$ to each advertising interval. The purpose of this delay is to avoid multiple subsequent colliding beacons. Further, in each advertising interval, BLE sends up to 3 consecutive beacons on 3 different channels (viz., channel 37, 38 and 39) in a row. The scanner toggles between these 3 channels for consecutive scan windows. Let the time within which these beacons are transmitted be $d_e$ time-units. Then, the effective advertising interval is increased by up to $\SI{10}{ms} + d_e$ time-units, as can be seen in Figure~\ref{fig:ble}. Here, $d_e$ accounts for the 3-channel discovery, whereas the $\SI{10}{ms}$ compensate for the largest possible random delay. Recall that SingleInt guarantees bounded worst-case latencies by requiring the distance between two consecutive packets to be less than or equal to $d_s - d_a$ time-units.
Hence, for compensating for the increased effective advertising interval, every scan window needs to be extended by $\SI{10}{ms} + d_e$ time-units beyond its optimal value. 

However, this extended scan window also increases the duty-cycle, which alters the optimal values for $T_a$, $T_s$ and $d_s$. Therefore, this overhead on $d_s$ needs to be accounted for in Equation~\ref{eq:etaDef}. In addition, the 3 beacons sent on different channels per advertising interval incur a certain overhead $o_a$ every $T_a$ time-units. This overhead is given by the time needed to transmit these additional beacons, plus the idle times between these transmissions weighted by the reduced power consumption compared to the transmission phases.
 With this, Equation~\ref{eq:etaDef} can be written as follows.
\begin{equation}
\label{eq:etaDefBle}
\eta = \frac{(d_s + \SI{10}{ms} + d_e)}{T_s} + \alpha \cdot \frac{d_a + o_a}{T_a}
\end{equation}
In Equation~\ref{eq:etaDefBle}, one could further artificially increase $T_a$ by the mean value of the random delay (i.e. $\SI{5}{ms}$), which we do not consider further because it only has a negligible impact. 
From this, $T_a$, $T_s$ and $d_s$ can be derived as described in Section~\ref{sec:PIkMProtocol}. 

With the extended scan window, the worst-case latencies of BLE configured using these values does not differ by more than $\SI{10}{ms}$ from those of ideal PI-based protocols configured using the same values for $T_a, T_s$ and (the non-extended) $d_s$.

\subsubsection{Remaining Advertising Modes}
The \textit{non-connectable undirected} advertising mode of BLE implies that a device receiving a beacon can neither establish a connection nor request additional data. We next study how the bidirectional modes of BLE, i.e., \textit{connectable} and \textit{scannable} advertising, can be configured using the SingleInt scheme.

In such modes, for each advertising beacon being sent, the transmitting device needs to listen to the channel for incoming responses $\SI{150}{\micro s}$ later. Hence, in addition to $o_a$, each beacon transmission duration is increased by an effective overhead of $o_{a2}$ time-units that accounts for this reception window, which needs to be accounted for in Equation~\ref{eq:etaDefBle}.
Note that blocking does not occur in such scenarios, since the receiving device only transmits a single packet after it has received a beacon from the transmitting device. Accounting for the above, the parameter values can be derived  as described in Section~\ref{sec:PIkMProtocol}.

\subsubsection{Adapting the MultiInt Scheme to BLE}
For parametrizations following the MultiInt scheme, in general, the same procedure as described above will lead to beneficial values of $k_c$ (and hence, $T_a$, $T_s$ and $d_s$) for BLE. However, the MultiInt scheme guarantees discovery within multiple instances of the scan interval. Recall that the distance of any scan window and its neighboring advertising beacon on the left is successively reduced by $\gamma$ time-units after every scan interval $T_s$. For each such reduction, $n = \lceil \nicefrac{T_s}{T_a} \rceil$ advertising intervals pass. Bounded discovery latencies can be guaranteed if $\gamma < d_s - d_a$, and since $\gamma$ could exceed $d_s - d_a$ because of the random delay, $d_s$ needs to be increased to compensate for this. With a random delay of up to $\SI{10}{ms}$ per advertising interval, the sum of these delays can add up to $n \cdot \SI{10}{ms}$ time-units after every $n$ intervals. This sum can easily exceed $T_a$ time-units and hence, a compensation by extending $d_s$ becomes very energy-consuming.

Therefore, the maximum random delay per interval needs to to be reduced to a reasonably low value, e.g.,  $\nicefrac{1}{n} \cdot \SI{10}{ms}$, compared to the value of $\SI{10}{ms}$ suggested by the BLE specification~\cite{bleSpec50}. Since this does not comply with the BLE specification, unlike the SingleInt scheme, the MultInt scheme cannot be implemented using most commercial BLE stacks, which typically do not support modifying the range of random values. However, if the random delay can be adjusted e.g., by using an open-source stack, the resulting implementation remains compatible with devices that use the random delay specified in the BLE standard. In particular, they can also be discovered by scanners that use proprietary stacks.

%% file: sections/evaluation_ii.tex
\section{Evaluation}
\label{sec:evaluation}
In this section, we evaluate the performance (i.e., worst-case latency, channel utilization and blocking probability) of the blocking-compensated version of the MultiInt scheme with $M=2$, as described in Section~\ref{sec:implementation} (viz., MultiInt-BC (M=2)). We compare its performance to those of multiple popular previously known protocols and evaluate its behavior in real-world implementations. 
Recall that our proposed parametrizations target networks with few devices discovering each other simultaneously. We therefore assume a symmetric scenario with two devices, for which the results of this comparison are valid. We further evaluate the predicted performance by real-world measurements on $560,000$ discovery procedures using two nRF51822~\cite{nrf51822:14} radios. Finally, we evaluate the performance of BLE configured using the SingleInt scheme.
\subsection{Protocols Considered}
%In the following, we present a selection of protocols to which we compare our proposed solution. 
%To maintain comparability, we do not consider any indirect approaches, e.g., \cite{zhang:12, julien:17, kindt:17a}, in which  information on future wakeup phases is transmitted along with the packets to accelerate the discovery.
We compare our proposed solution to the following ones.
\begin{asparaitem}
	\item \textbf{Disco}~\cite{dutta:08} is used as the baseline for comparisons in most related work. Though Disco can only function properly with two prime numbers $p_1$ and $p_2$, the equations describing the performance in \cite{dutta:08} remain valid if we assume $p_1 = p_2$ and also allow non-prime numbers, as proposed in ~\cite{dutta:08}.
	This allows for a comparison of all duty-cycles, including those that cannot be realized in practice due to the lack of corresponding prime numbers.
	Note that assuming $p_1 = p_2$ in the equation describing Disco's performance is not equivalent to actually implementing Disco using the same prime numbers, which would negatively affect its performance. 
	\item \textbf{U-Connect}~\cite{Kandhalu:10} allows for extremely short slot lengths of $\SI{250}{\micro s}$, which leads to a high performance.
	\item \textbf{Searchlight-Striped}~\cite{bakht:12} achieves low latencies, while being capable of realizing a large set of duty-cycles.
	\item \textbf{Optimal Diffcodes}~\cite{meng:14} have been proven to achieve the lowest worst-case latencies (in terms of slots) that any slotted protocol could guarantee~\cite{zheng:03, kindt:19}. Despite only a very limited set of duty-cycles can be realized, we consider their theoretical performance for all duty-cycles.
	\item \textbf{G-Nihao}~\cite{qiu:16} defines listen-only and transmit-only slots. It provides a parameter $\gamma$ to adjust the number of beacons per period, but its optimal value is not clear. We therefore assume $\gamma = 2$, as also assumed in \cite{qiu:16}.
\end{asparaitem}

The corresponding worst-case latencies, based on \cite{Kandhalu:10, bakht:12, meng:14, qiu:16}, are given by Table \ref{tab:discovery-latency-table}. Here, $d_{sl}$ is the slot length.
Whenever necessary, the Equations from the literature have been rearranged to account for the assumptions described above and to bring them into the same form. We use these latency-duty-cycle relations to asses the worst-case performance of the previously known protocols in our comparison.
\begin{table}
	\begin{center}
		{\renewcommand{\arraystretch}{1.3}
			\begin{tabular}{cc}
				\hline Protocol & $d_m(\eta)$ \\ 
				\hline Disco & $\frac{4}{\eta^2} d_{sl}$ \\ 
				U-Connect & $(\sqrt{\frac{1}{2 \eta} + \frac{9}{16 \eta^2}} + \frac{3}{4 \eta})^2 d_{sl}$ \\ 
				Searchlight & $\left\lceil \frac{\left\lfloor\frac{1}{\eta}\right\rfloor}{2}\right\rceil d_{sl}$\\
				Optimal Diffcodes & $\frac{1}{2 \eta^2} d_{sl}$\\
				G-Nihao  & $ \left(\frac{d_{sl} + d_a \gamma}{2 \gamma \eta d_{sl}} + \sqrt{\frac{d_{sl} + d_a \gamma}{2 \gamma \eta d_{sl}}- \frac{d_a}{d_{sl}}}\right)^2 \gamma$ \\
				\hline 
			\end{tabular} 
		}
	\end{center}
	\caption{Worst-case discovery latencies of slotted protocols.}
	\label{tab:discovery-latency-table}
\end{table}

\subsection{Protocol Parameters}
The performance of all \ac{ND} protocols depends on the packet- and slot lengths, for which we derive reasonable values next.
\subsubsection{Beacon Transmission Duration ${d_a}$}
Slotted and PI-based protocols perform best for short beacon lengths. We therefore assume a length of $4$ bytes, which consists of a 1-byte preamble for synchronization and a 3-byte timestamp to schedule a later data exchange. We assume a nRF51822-radio \cite{nrf51822:14} with a bitrate of $\SI{1}{Mbit/s}$, which leads to a transmission duration $d_a$ of $\SI{32}{\micro s}$.

\subsubsection{Slot Length ${d_{sl}}$}
Whereas PI-based protocols guarantee discovery within a certain amount of time, slotted protocols guarantee discovery within a worst-case number of slots.
Shorter slot lengths lead to lower worst-case latencies and hence a higher performance for the same duty-cycle.
A comparison using the slot lengths assumed in the literature would not
be fair, since they are based on ``good guesses'' and have not been chosen to systematically minimize the discovery latencies or failure probabilities. A systematic reduction of the slot length has not yet been studied. Therefore, we need to identify the slot lengths for which different slotted protocols achieve the same properties as PI-based ones. As already mentioned, for the largest duty-cycle considered, implementations following the MultiInt-BC scheme with $M=2$ will fail (and hence prevent a successful discovery) in around $\SI{0.19}{\percent}$ of all cases due to beacon collisions and blocking. For slotted protocols, this rate grows for decreasing slot lengths, and we in the following identify the slot lengths that lead to the same fraction of failed discoveries as in MultiInt-BC.

For a successful discovery, two slots from two devices have to overlap in time. This also implies that beacons from two devices come into temporal vicinity, which makes slotted schemes prone to collisions even when their channel utilization is low. Consider a slot in which a beacon is transmitted at its beginning and end, whereas the devices listen to the channel in between. Two such slots from different devices can only overlap, if the difference of their starting times lies within $[-d_{sl}, d_{sl}]$ time-units. For certain offsets within this range, discovery will be prevented due to beacon collisions, e.g., for the offset $0$. In addition, the radio has to switch from reception to transmission and vice-versa, which blocks $d_{rt}$ and $d_{tr}$ time-units in each slot. One can compute the probability of failed discoveries by integrating over all offsets that lead to failures and dividing them by the range of offsets with overlapping slots, which leads to the following failure probability for 2 devices discovering each other:
\begin{equation}
\label{eq:pblkdisco}
P_{blk,disco} = \frac{2 \cdot (3 d_a + d_{rt} + d_{tr})}{2 d_{sl}}.
\vspace*{1em}
\end{equation}
This slot design is actually used by Disco \cite{dutta:08}, for which Equation \ref{eq:pblkdisco} gives the probability of failed discoveries. In contrast, Searchlight~\cite{bakht:12} and optimal difference codes~\cite{meng:14} define overflowing slots, in which at least one beacon is sent slightly outside of the slot boundaries. 
Under the assumption that one beacon transmission and turnaround phase lie within the slot, whereas another turnaround phase and beacon transmission lie outside of the slot boundaries, a reduced probability of 
\begin{equation}
\label{eq:pblkoverflow}
P_{blk,overflow} = \frac{2 d_a + d_{tr}}{d_{sl}}
\end{equation}
can be achieved for these protocols.
In G-Nihao, there are always $m$ consecutive listen-only slots, which can be regarded as a large, contiguous reception slot. Since beacons are also sent with a period of $m$ slots, always one beacon transmission duration $d_a$ and a pair of turnaround phases $d_{tr}$ and $d_{rt}$ lie within these $m$ slots. This leads to the following probability:
\begin{equation}
\label{eq:pblknihao}
P_{blk,nihao} = \frac{d_{rt} + d_{tr} + 2 d_a}{m \cdot d_{sl}}
\end{equation}

As already mentioned, the MultiInt-BC scheme with $M=2$ achieves a blocking probability of around $\SI{0.19}{\percent}$ for two devices and for the least beneficial duty-cycle considered (i.e., $\eta = \SI{1.55}{\percent}$). 
Therefore, in this comparison, we scale the slot length of each slotted protocol, such that a maximum blocking probability of
$\SI{0.19}{\percent}$ for two devices and for the maximum duty-cycle considered (i.e., $\SI{1.55 }{\percent}$) is reached in each of them. This leads to the following slot lengths. Disco: $\SI{197.9}{\milli s}$; Searchlight and Diffcodes: $\SI{107.4}{\milli s}$; G-Nihao: $\SI{5.5}{\milli s}$. 
Since U-Connect defines special receive-only and transmit-only slots, we here assume $d_{sl} = \SI{250}{\micro s}$, as has been done in \cite{Kandhalu:10}\footnote{Therefore, the comparison with U-Connect does not follow our rationale of equivalent blocking probabilities.}.

\subsection{Worst-Case Latencies}
\label{sec:discovery_latencies}
\begin{figure}[bt]
	\centering
	\includegraphics[width=\linewidth]{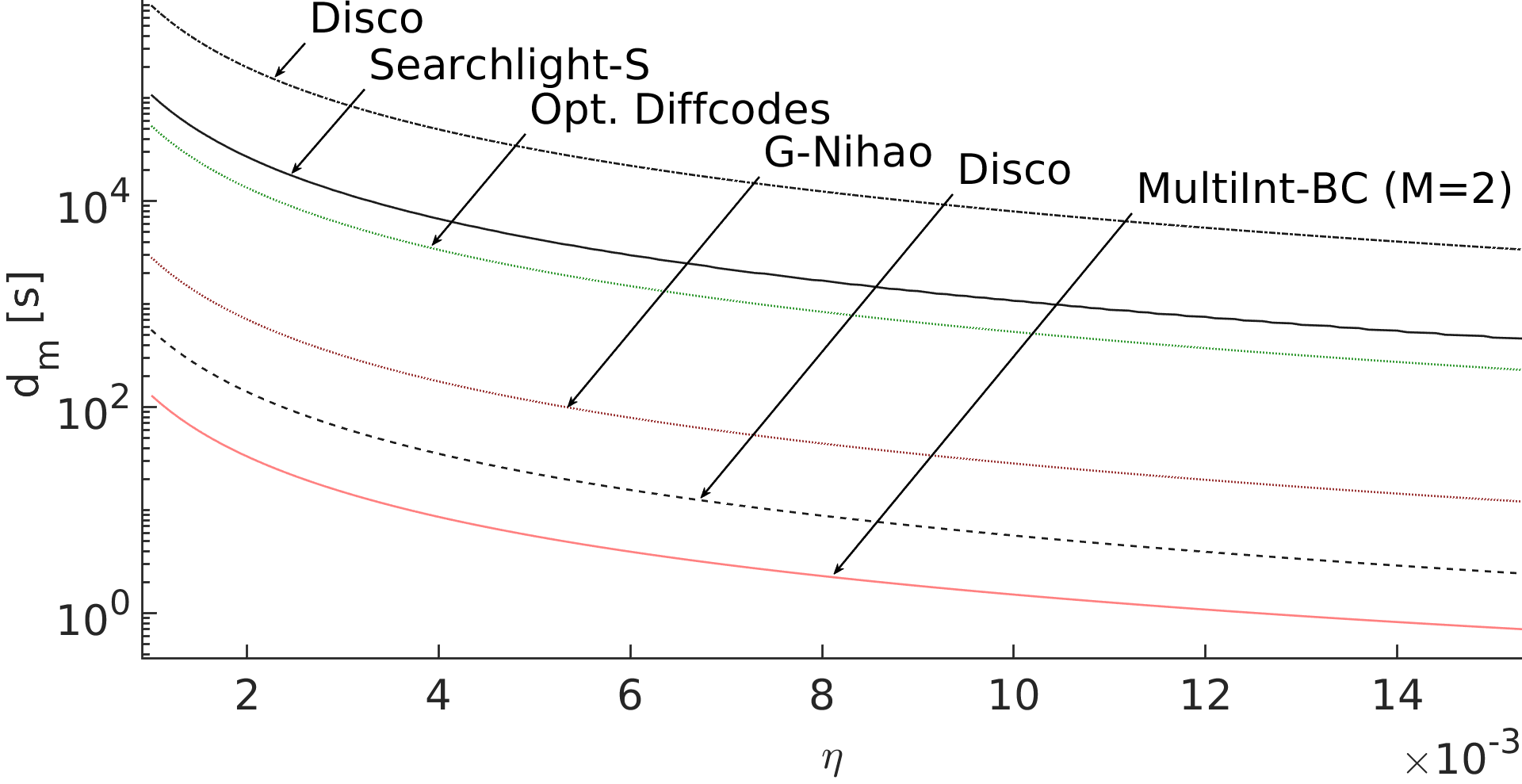}
	\caption{Worst case latencies of slotted protocols and the MulitInt-BC scheme.}
	\label{fig:performance_pi_protocols_vs_slotted} 
\end{figure}

In this Section, we evaluate the computed worst-case discovery latencies of the MultiInt-BC scheme with $M=2$, as described in Section~\ref{sec:implementation}. Further, we compare them to the latencies of previously known protocols, as depicted in Figure~\ref{fig:performance_pi_protocols_vs_slotted}. Note that if a protocol provides a shorter worst-case latency than another protocol for a given duty-cycle, this also implies that this device consumes less energy when guaranteeing the same worst-case latency.
	\footnotetext[2]{Static slot length and hence constant blocking probability.}
As can be seen, the slotted protocols Disco, Searchlight and Diffcodes have the highest worst-case latencies. U-Connect achieves lower worst-case latencies due to its separate listen-only and transmit-only slots. The pseudo-slotted protocol Nihao has a significantly lower worst-case latency, but does not reach the performance of the MultiInt-BC scheme, which provides the lowest worst-case latencies for all duty-cycles.
Table \ref{tab:pi_vs_slotted_results} shows the gains over slotted protocols, defined as $G = \frac{d_{m,protocol}}{d_{m,MultiInt-BC}}$. In particular, we have considered the maximum gains $G_m$ and the mean gains $\overline{G}$ over the entire range of duty-cycles considered.
For example, for the most beneficial duty-cycle of $\SI{0.2}{\percent}$, in the worst-case, U-Connect would take $4.4\times$ longer than the MultiInt-BC scheme for discovering a neighbor. On the average over all duty-cycles considered, U-Connect would take $4.1\times$ longer. In addition to the results for a maximum rate of failed discoveries of $\SI{0.19}{\percent}$, Table~\ref{tab:pi_vs_slotted_results} shows the results for a rate of $\SI{3}{\percent}$ for the highest duty-cycle. This rate is achieved by slotted protocols for two devices by adjusting the slot length accordingly, whereas the MultiInt-BC scheme fails with this rate when 3 devices come into range. As can be seen, there are still significant gains.
These results suggest that the classical slotted protocols Disco, Searchlight and Diffcodes, achieve larger worst-case latencies compared to PI-based ones. Further, PI-based ND protocols can achieve significantly lower worst-case latencies by decoupling reception and transmission.
Some configurations of the G-Nihao protocol~\cite{qiu:16} might potentially result in a similar sequence of packets and reception windows as the SingleInt scheme, if an optimal value for its parameter $\gamma$, which determines its channel utilization, could be found. The difference in performance in symmetric scenarios stems from the lower blocking probability of MultiInt-BC, as well as from the unknown optimal value of Nihao's $\gamma$-parameter.

\begin{table}
	\begin{center}
		{\renewcommand{\arraystretch}{1.0}
			\begin{tabular}{|c|cc|cc|}
				\hline & \multicolumn{2}{c|}{$P_{blk} = \SI{0.19}{\percent}$} & \multicolumn{2}{c|}{$P_{blk} = \SI{3}{\percent}$}\\
				& $G_m$ & $\overline{G}$ & $G_m$ & $\overline{G}$\\
				\hline
				Disco & 6119.1 & 5663.9 & 387.5 & 358.7\\ 
				Searchlight-S & 830.0 & 768.1 &52.6 & 48.6  \\ 
				Opt. DiffCodes & 415.5 &384.6 & 26.8 & 24.8 \\ 
				%Lightning & 1.6 & 1.6 \\ 
				G-Nihao & 22.0 & 20.3 & 1.7 & 1.6  \\ 
				U-Connect\footnotemark & 4.4& 4.1&4.4 & 4.1 \\ 
				\hline 
			\end{tabular} 
		}
		\caption{Maximum ($G_m$) and mean ($\overline{G}$) gains of worst case discovery latencies over slotted protocols achieved by the \textit{MultiInt-BC} scheme over all duty-cycles considered.}
		\label{tab:pi_vs_slotted_results}
		
	\end{center}
\end{table}
\subsection{Average-Case Behavior}
In what follows, we compare the mean latencies of the MultiInt-BC scheme to those of previously known protocols. Since the literature does not provide equations on the mean latencies, we have implemented simulation models of all previously known protocols under consideration. 
For this comparison only, we assume a duty-cycle of $\SI{5}{\percent}$, because it can at least be approximated closely by all protocols under consideration (e.g., using the prime numbers 37 and 43 for Disco). 
\begin{figure}[t!]
	\centering
	\includegraphics[width=\linewidth]{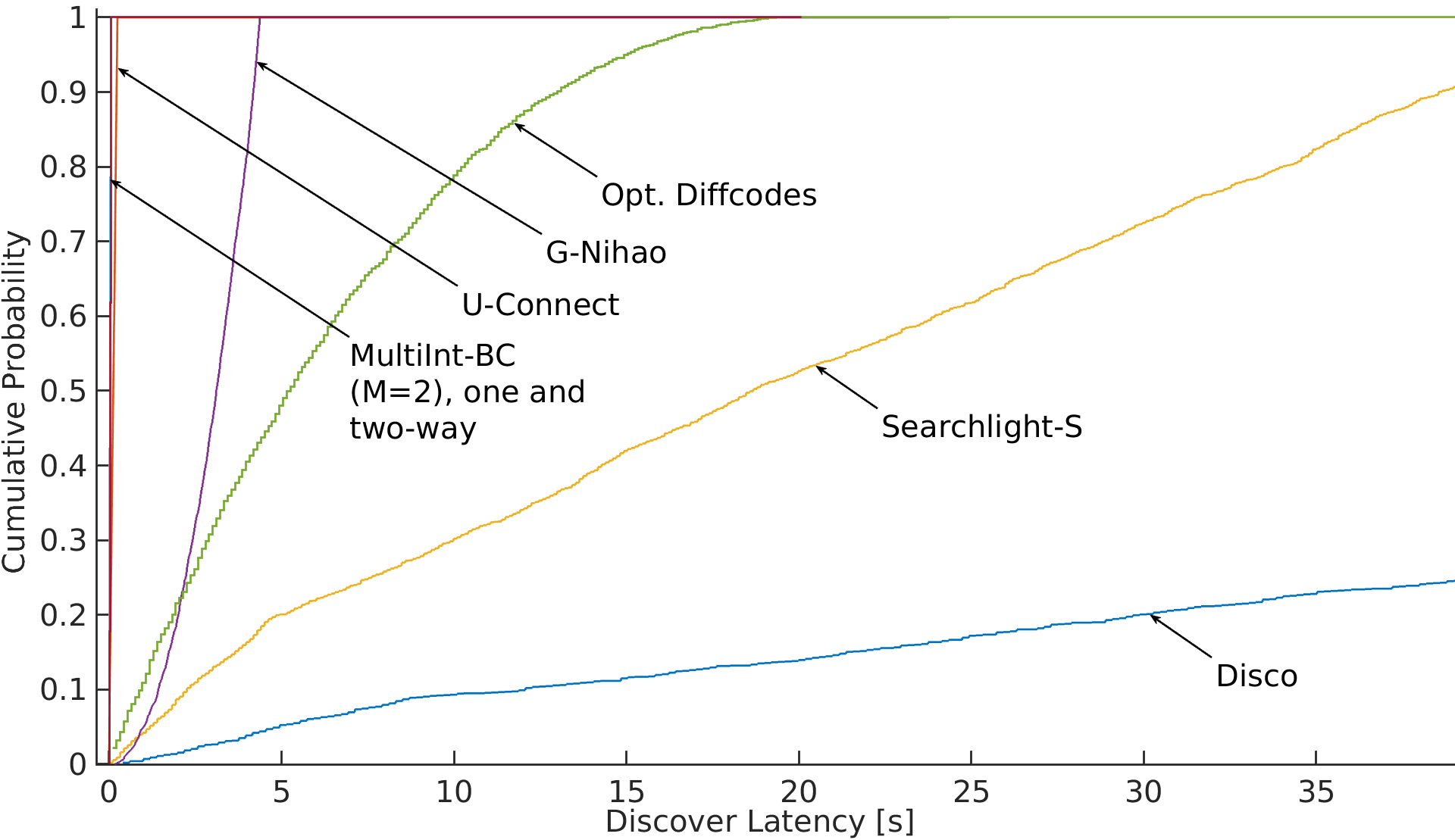}
	\caption{Computed/Simulated CDFs for a duty-cycle of $\SI{5}{\percent}$ under the absence of collisions.}
	\label{fig:PIkM-CDF} 
\end{figure}
Figure~\ref{fig:PIkM-CDF} shows the cumulative distribution functions (CDFs) of our proposed solution and the previously known ones. Here, we have assumed that no collisions occur for all considered protocols.
We have neglected the small impact on the timing behavior caused by the blocking countermeasures described in Section~\ref{sec:implementation}, but have accounted for the duty-cycle increase (which is, as we have already described, non-negligible).
For the MultiInt-BC scheme with $M=2$, both the CDF-curves for one-way discovery (i.e., a device A receives a packet from device B) and for two-way discovery (i.e., device A receives a packet from device B and vice-versa) are depicted, but both curves lie in such a close proximity that hardly any difference is visible. The remaining depicted CDF curves represent two-way discoveries. 
For a duty-cycle of $\SI{5}{\percent}$, the mean latencies are as follows:
Disco: $\SI{104.20}{s}$; Searchlight-Striped: $\SI{19.71}{s}$; Difference Sets: $\SI{6.16}{s}$; G-Nihao: $\SI{2.94}{s}$; U-Connect: $\SI{0.12}{s}$; MultiInt-BC, two-way $\SI{0.04}{s}$; MultiInt-BC, one-way: $\SI{0.03}{s}$.

\subsection{Channel Utilization and Collisions}
\begin{figure}[tb]
	\centering
	\includegraphics[width=\linewidth]{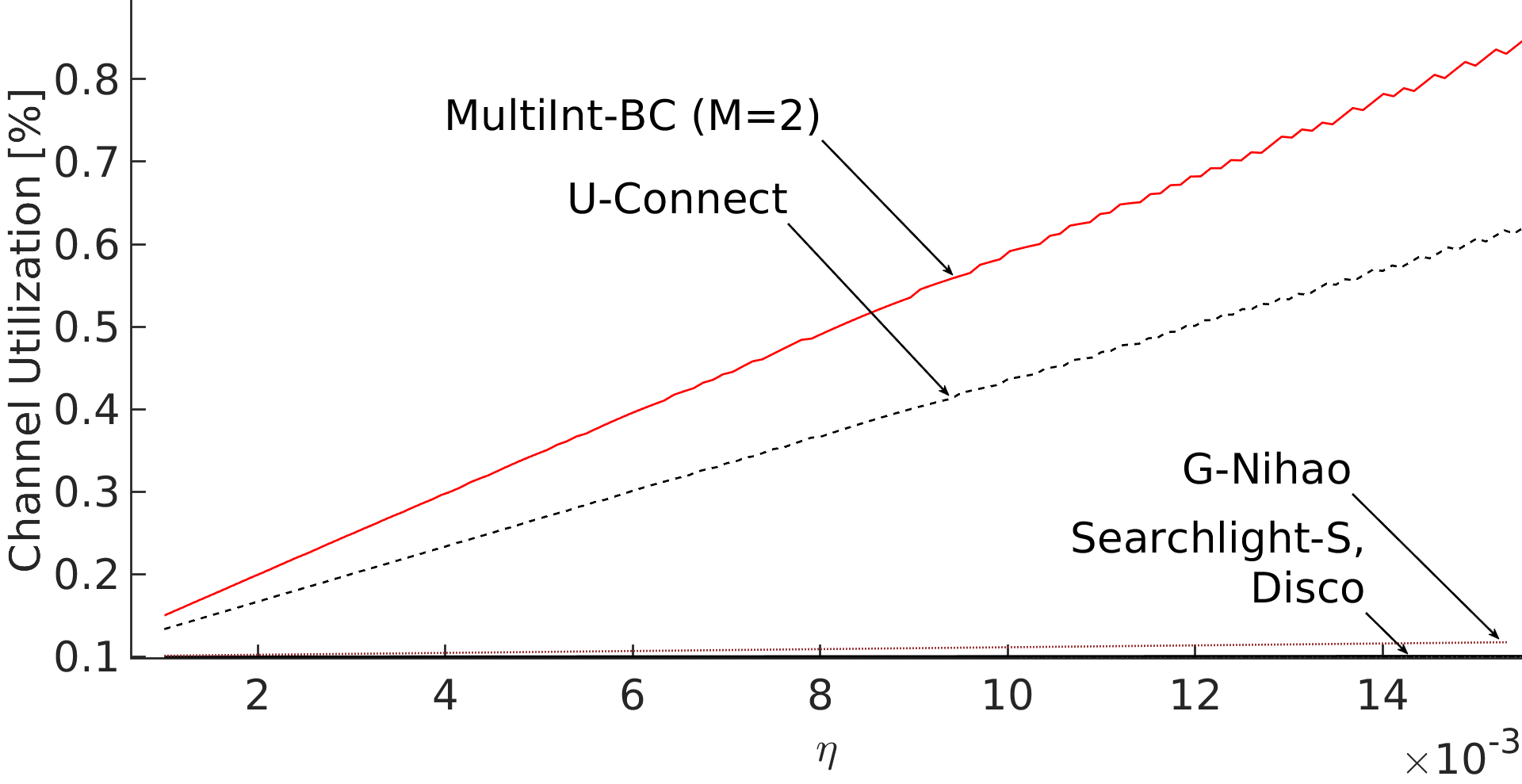}
	\caption{Comparison of channel utilization.}
	\label{fig:channel_utilization_pi_protocols_vs_slotted} 
\end{figure}

A comparison of the computed channel utilization is depicted in Figure~\ref{fig:channel_utilization_pi_protocols_vs_slotted}. The MultiInt-BC scheme adjusts the transmission rate for optimal latency-duty-cycle relations, which leads to an increased channel utilization. Recall that our proposed schemes target scenarios with few nodes being in discovery mode simultaneously, and we in the following establish that the resulting collision rates remain low in the scenarios considered. 
\label{seq:collisions}
As already mentioned, for 2 devices, the blocking mitigation techniques described in Section \ref{sec:implementation} lead to a probability of blockage and collisions of up to $\SI{0.19}{\percent}$. This is achieved by keeping all reception phases free of packet transmissions, thereby preventing failures due to collisions. When more than two devices are in range, collisions will occur regardless of this. Since the offsets of packets from different devices are usually distributed uniformly, their collision probabilities are exponentially distributed (cf. \cite{liu:13} for details).
Therefore, starting from $nDevices = 3$ devices, the discovery procedure of each device will collide with a probability of 
\begin{equation}
\label{eq:piK2collisions}
p_{col} = 1 - e^{-2 (nDevices - 1) \cdot (\frac{d_a}{T_a} + 2 \frac{d_a}{T_s})}.
\end{equation}
For 3 devices, the collision probability is around $\SI{0.5}{\percent}$ for $\eta = \SI{0.2}{\percent}$ and around $\SI{3}{\percent}$ for $\eta = \SI{1.55}{\percent}$, as we had assumed in our comparison to slotted protocols. For 10 devices, the collision probability is around $\SI{2}{\percent}$ for $\eta = \SI{0.2}{\percent}$ and reaches almost $\SI{13}{\percent}$ for $\eta = \SI{1.55}{\percent}$. 
From these results, we can conclude that PI-based protocols parametrized using our proposed schemes perform optimally in the unidirectional case and essentially optimally (since $d_m$ is near-minimal and only a negligible number of discovery attempts fail) in the symmetric case for two devices. For more than two devices, the performance gracefully decreases, while remaining feasible for networks with up to 10 devices being in discovery mode simultaneously. The failure probability in one-way scenarios scales with the number of senders, while the number of passive receivers does not influence the failure rate.

\subsection{Experimental Latency Measurements}
\begin{figure}[hbt]
	\begin{subfigure}
		\centering
		\includegraphics[width=\linewidth]{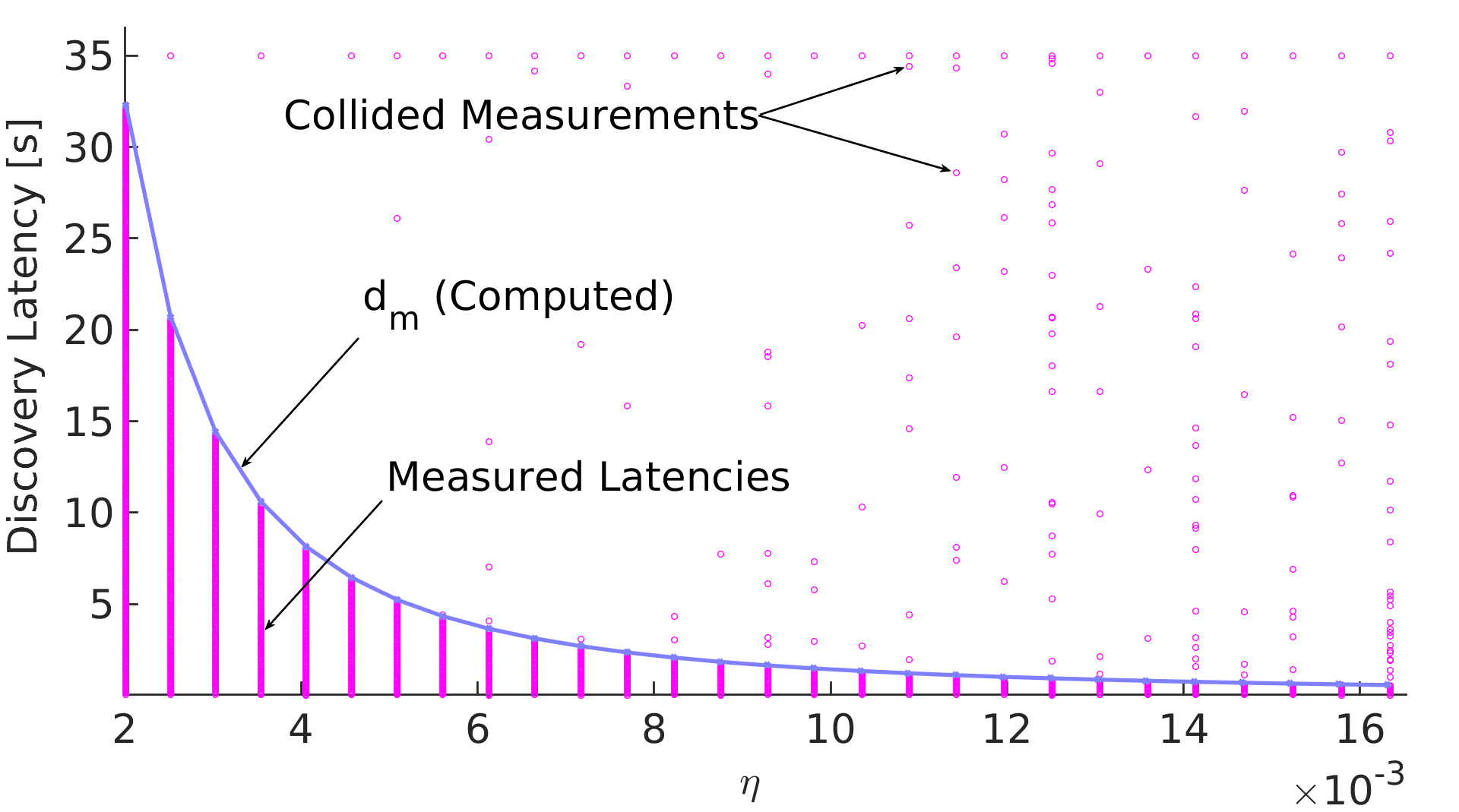}
		\caption{Measured discovery latencies (scattered points) and computed upper latency bound (solid line).}
		\label{fig:measurement_results} 
	\end{subfigure}
	\begin{subfigure}
		\centering
		\includegraphics[width=\linewidth]{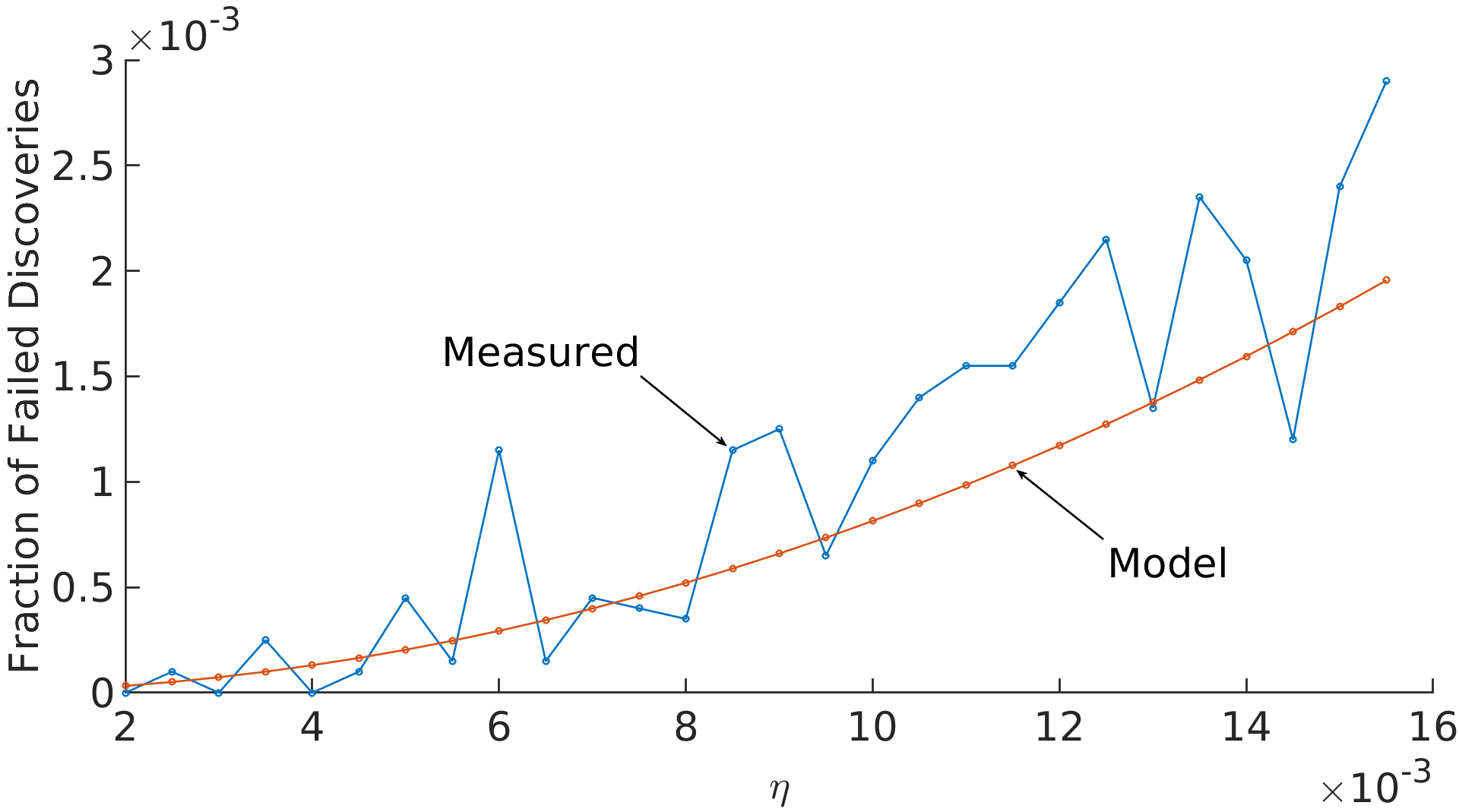}
		\caption{Predicted and measured fraction of failed discoveries.}
		\label{fig:piKM_blocking} 
	\end{subfigure}
	
\end{figure}

To demonstrate that our proposed \ac{ND} solution can be realized in practice, we have implemented the MultiInt-BC scheme, as described in Section \ref{sec:implementation}, on two wireless radios. Based on the open-source BLE stack Blessed \cite{blessed:15}, we have created a custom firmware for two nRF51822-radios. The radios were connected to an ARM Cortex M4 microcontroller via UART. The purpose of this microcontroller was to start and stop the radios, to send them the appropriate parametrizations, to obtain reports on the received packets and to measure the discovery latencies.
Both radios have been located in close proximity within an RF-shielded, anechoic box. In multiple runs, the radios repeatedly discovered each other using the MultiInt-BC scheme. After each discovery attempt, the devices have been desynchronized by a random waiting time. In each measurement, both devices were started such that every two neighboring scan windows on two devices had a random time offset within $[0, T_s]$ between each other. Similarly, any pair of beacons on two devices had a random offset within $[0, T_a]$ time-units between each other. 
%Since both devices have been permanently within their range of reception during our experiments, we have implemented a mechanism to virtually bring the two radios in or out of range. Whenever both radios have started advertising and scanning, the ARM microcontroller controlling both radios has waited a random amount of time between $0$ and $T_s$ plus a certain safety margin. This waiting time represents the phase in which the radios were not in range, and all receptions that took place within it were ignored. After this waiting time, the devices were considered to be in range and received packets were considered as successful discoveries. 
After both devices had either discovered each other, or after a timeout of 35 seconds (which exceeded $d_m$ for all duty-cycles) was reached, the radios were stopped, the measured latencies were logged on a laptop and the next measurement round was initiated. For each duty-cycle, the experiment was repeated $10,000$ times, leading to $20,000$ one-way discoveries. We have considered 28 different duty-cycles between $\SI{0.2}{\percent}$ to $\SI{1.55}{\percent}$, which resulted in $560,000$ measured discovery procedures.

The measured discovery-latencies, together with the computed upper limit $d_m(\eta)$, are shown in Figure \ref{fig:measurement_results}. Each scattered point represents a measured latency of one discovery procedure, whereas the solid line depicts the upper bound predicted by our theory. As can be seen, the measured latencies always lie below this bound, except for a few collided attempts. 

Figure \ref{fig:piKM_blocking} shows the measured fraction of discoveries that have exceeded the predicted worst-case bound $d_m$ by more than $\SI{1}{\percent}$. Deviations below $\SI{1}{\percent}$ have been considered as measurement inaccuracies. In addition, the predicted fraction of failed discoveries from Equation \ref{eq:PIk2blockage} is shown. As can be seen, the measurements match the predicted values well. For a duty-cycle of $\SI{0.2}{\percent}$, all $20,000$ discoveries were successful, whereas the maximum number of failures was $58$ for $\eta=\SI{1.55}{\percent}$. This corresponds to the failure rate of $\SI{0.29}{\percent}$ depicted in Figure~\ref{fig:piKM_blocking}. These results show that our proposed MultiInt-BC scheme reaches the predicted latencies in practice, while also offering very low failure probabilities.

\subsection{Performance of BLE}
\label{sec:ble_performance}
In Section~\ref{sec:ble}, we have described how our proposed parametrization framework can be used to optimize BLE. Recall that BLE requires that a random delay is added to each instance of $T_a$. In addition, a reception phase after each transmission is required in the case of bi-directional discovery. This affects its latency-duty-cycle performance. In the following, we evaluate how the performance of BLE configured using the SingleInt-BLE scheme compares to an ideal PI-based protocol configured using the SingleInt scheme (i.e., SingleInt without any overheads of and modifications for BLE). The values for this comparison have been obtained from computations. 

For this evaluation, we assume the following overheads for BLE: $o_a = \SI{619}{\micro s}$, $o_{a2} = \SI{143}{\micro s}$, $o_s = \SI{11}{ms}$. They result from the following assumptions:
\begin{asparaitem}
	\item We assume a packet length of 30 bytes, which is a realistic value for BLE (e.g., for a location beacon).
	\item We assume that any two consecutive packet transmissions on two different channels are spaced from each other by $\SI{150}{us}$. Further, we assume that the radio consumes the same power during the time between two consecutive transmissions as for switching from transmission to reception.
	\item We assume that the power consumption for transmission is identical to that for reception.
	\item We assume the following values from the literature~\cite{kindt:13} for a BLE radio: Idle-listening (i.e., the short listening phase after transmitting a packet) takes $\SI{74}{\micro s}$ and the quotient of the power consumption for transmission over that for switching from reception to transmission has a value of $0.46$.
	\item  We study a range of duty-cycles between $\SI{2.15}{\percent}$ and $\SI{10}{\percent}$. The range of considered duty-cycles for SingleInt-BLE needs to be larger than for the unmodified SingleInt scheme, since BLE uses significantly larger beacon lengths, leading to increased duty-cycles for reaching the same worst-case latency. 
\end{asparaitem}

\begin{figure}[tb]
	\centering
	\includegraphics[width=\linewidth]{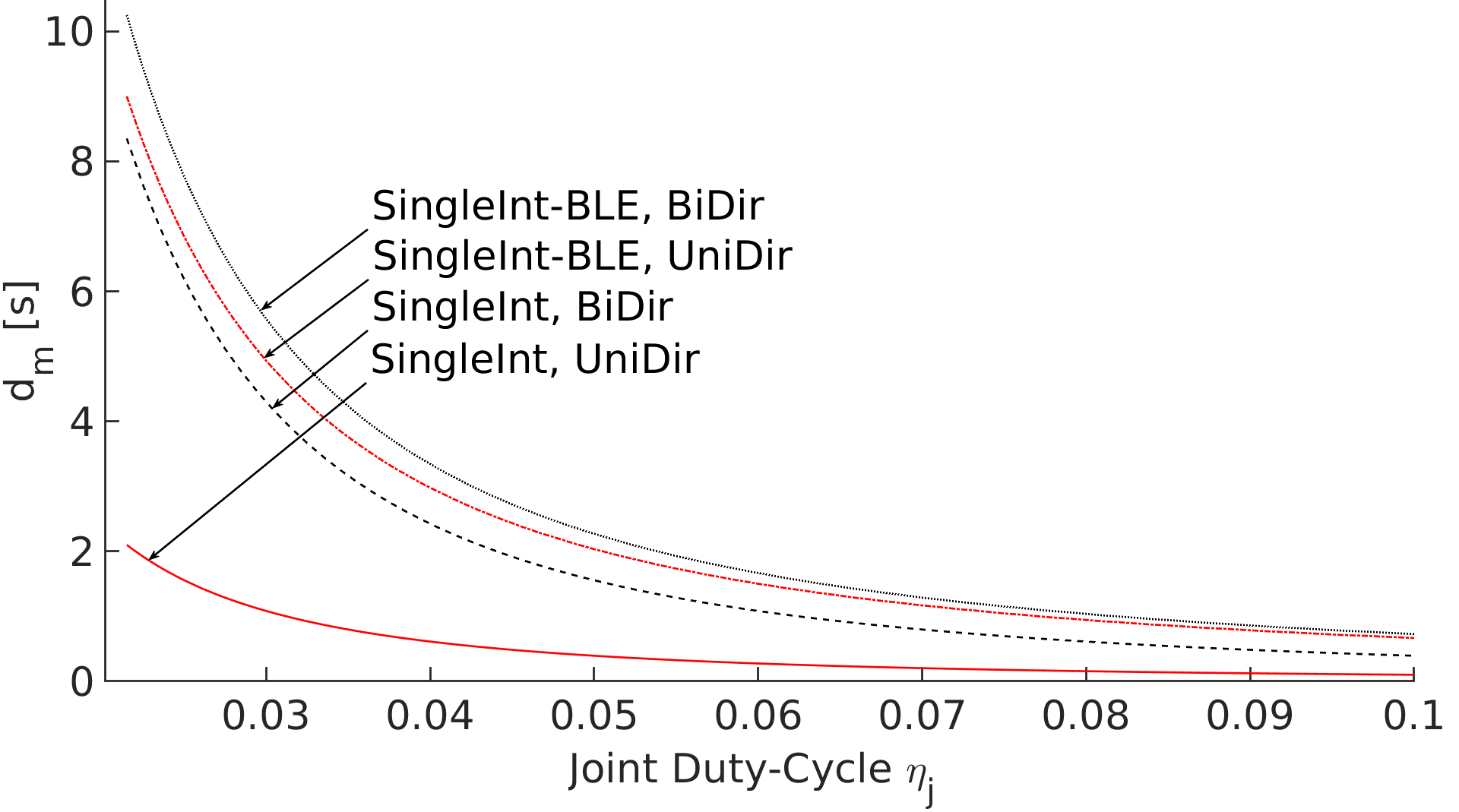}
	\caption{Performance of BLE configured using SingleInt.}
	\label{fig:ble_performance} 
\end{figure}
Figure~\ref{fig:ble_performance} depicts the worst-case latencies of BLE configured using the SingleInt-BLE scheme, both for the unidirectional and bidirectional advertising modes. In addition, the worst-case latencies of an unmodified PI protocol configured according to the SingleInt scheme without the modifications for BLE are shown. The depicted duty-cycle $\eta_j$ is the joint duty-cycle of both devices, i.e., the sum of the duty-cycles of two devices that carry out the discovery procedure.
For unidirectional discovery, one can see that the worst-case latencies of SingleInt-BLE are increased compared to the latencies obtained by protocols applying the original SingleInt scheme. On the average, for unidirectional discovery, SingleInt-BLE has a $5.5 \times$ larger worst-case latency than the original SingleInt scheme. For bidirectional discovery, SingleInt-BLE on the average has a $1.5 \times$ larger worst-case latency.
\begin{table}
	\centering
	\begin{tabular}{|c|c|c|}
		\hline 
		Value & BLE, UniDir & BLE, BiDir \\ 
		\hline 
		$T_a$ & $\SI{23}{ms} - \SI{88}{ms}$ &  $\SI{27}{ms} - \SI{101}{ms}$\\ 
		$T_s$ & $\SI{655}{ms} - \SI{8.990}{s}$ &  $\SI{715}{ms} - \SI{10.241}{s}$\\ 
		$d_s$ & $\SI{35}{ms} - \SI{99}{ms}$ &  $\SI{38}{ms} - \SI{112}{ms}$\\ 
		\hline 
	\end{tabular} 
	\caption{Parameter values chosen by SingleInt-BLE.}
	\label{tab:bleIntervalLengths}
\end{table}
Table~\ref{tab:bleIntervalLengths} depicts the range of interval lengths chosen by the SingleInt-BLE scheme in the considered range of duty-cycles. 
As can be seen, they are considerably larger than for the unmodified SingleInt scheme. This range of values complies to the \ac{BLE} standard and is supported by typical commercial BLE stacks. E.g., the Nordic S110 SoftDevice supports values of $T_a$ between $\SI{20}{ms}$ and $\SI{10.24}{s}$ and values of $T_s$ and $d_s$ between $\SI{2.5}{ms}$ and $\SI{10.24}{s}$~\cite{nordicSoftDev:19}.

%% file: sections/concluding_remarks.tex
\vspace*{-0.5em}
\section{Concluding Remarks}
\label{sec:conclusion}
We have introduced a parametrization scheme for slotless, PI-based \ac{ND} protocols. Since such protocols can make use of more degrees of freedom than slotted ones, they can optimize their beacon transmission rate and achieve significantly lower discovery latencies in scenarios with few devices discovering each other simultaneously. In addition, unlike most previously proposed deterministic protocols, PI-based ones can realize practically every specified duty-cycle. Therefore, they are a practical choice for many personal area networks and IoT scenarios. One variant of our scheme can also be used for parameterizing BLE, which makes it highly relevant for practical applications.
With a PI-based protocol configured according to our proposed scheme, a protocol with provably optimal performance is available. It performs optimally for unidirectional discovery between one sender and one receiver, and near-optimally for bi-directional symmetric discovery between two devices. For larger numbers of devices discovering each other, collisions will play an increasing role, and hence an increasing number of discovery attempts will fail. For such scenarios, the development of an optimal \ac{ND} protocols remains open for future research.

%% file: sections/app_constraints.tex
\section{Deviation of Constraints of the SingleInt Scheme}
\label{sec:app_constraints}
In this appendix, we give a detailed derivation of the constraints of the SingleInt scheme, which have been described in Section~\ref{sec:PI0M_constraints}.

Given a lower limit $d_{s,m}$ of the scan window that the radio hardware can realize, we always enforce $d_s(\eta) \geq d_{s,m}$.
Using Equation~\ref{eq:0M_constraint_inequality}, we can write this inequality as
\begin{equation}
\label{eq:dsbyEta0Mrepeat}
d_s(\eta) = \frac{d_a (M+1)(\eta + 1)}{(\eta(M+1) - 1)} \geq d_{s,m}
\end{equation}
When solving this inequality, two cases can occur:\\
\\
1) $d_a (M+1)(\eta + 1) \geq d_{s,m} \cdot (\eta (M+1) - 1)$ and $\eta(M+1) - 1$ > 0\\

\noindent This condition implies that $M > \nicefrac{1}{\eta} - 1$ and $(M+1)\cdot(\eta(d_a - d_{s,m})+ d_a) \geq - d_{s,m}$. From this directly follows
that if $\eta (d_a - d_{s,m}) + d_a >0$ and hence, $\eta < \nicefrac{d_a}{d_{s,m} - d_a}$, then it is
\begin{equation}
M \geq \frac{d_{s,m}(\eta - 1) - d_a \cdot (\eta + 1)}{d_a \cdot (\eta + 1) - \eta d_{s,m}}
\end{equation}
This is already fulfilled by $M > \nicefrac{1}{\eta} - 1$, which we have required above.
Further, if $\eta \cdot (d_a - d_{s,m}) + d_a < 0$ and hence, $\eta > \nicefrac{d_a}{d_{s,m} - d_a}$ then ,

\begin{equation}
M \leq \frac{d_{s,m}(\eta - 1) - d_a \cdot (\eta + 1)}{d_a \cdot (\eta + 1) - \eta d_{s,m}}
\end{equation}
\\
2) $d_a (M+1)(\eta + 1) \leq d_{s,m} \cdot (\eta (M+1) - 1)$ and $\eta(M+1) - 1 < 0$\\

This case is not feasible, since $d_s(\eta)$ from Equation~\ref{eq:dsbyEta0Mrepeat} would become negative.

In summary, we require that
\begin{equation}
M > \frac{1}{\eta} - 1
\end{equation}
If $\eta > \nicefrac{d_a}{d_{s,m} - d_a}$, we further require 
\begin{equation}
M \leq \frac{d_{s,m}(\eta - 1) - d_a \cdot (\eta + 1)}{d_a \cdot (\eta + 1) - \eta d_{s,m}}
\end{equation}

%% file: paper_ieee.bbl
\begin{thebibliography}{10}
\providecommand{\url}[1]{#1}
\csname url@samestyle\endcsname
\providecommand{\newblock}{\relax}
\providecommand{\bibinfo}[2]{#2}
\providecommand{\BIBentrySTDinterwordspacing}{\spaceskip=0pt\relax}
\providecommand{\BIBentryALTinterwordstretchfactor}{4}
\providecommand{\BIBentryALTinterwordspacing}{\spaceskip=\fontdimen2\font plus
\BIBentryALTinterwordstretchfactor\fontdimen3\font minus
  \fontdimen4\font\relax}
\providecommand{\BIBforeignlanguage}[2]{{%
\expandafter\ifx\csname l@#1\endcsname\relax
\typeout{** WARNING: IEEEtran.bst: No hyphenation pattern has been}%
\typeout{** loaded for the language `#1'. Using the pattern for}%
\typeout{** the default language instead.}%
\else
\language=\csname l@#1\endcsname
\fi
#2}}
\providecommand{\BIBdecl}{\relax}
\BIBdecl

\bibitem{dutta:08}
P.~Dutta and D.~Culler, ``Practical asynchronous neighbor discovery and
  rendezvous for mobile sensing applications,'' in \emph{ACM Conference on
  Embedded Network Sensor Systems (SenSys)}, 2008.

\bibitem{bakht:12}
M.~Bakht, M.~Trower, and R.~Kravets, ``Searchlight: won't you be my neighbor?''
  in \emph{Annual International Conference on Mobile Computing and Networking
  ({MOBICOM})}, 2012.

\bibitem{Kandhalu:10}
A.~Kandhalu, K.~Lakshmanan, and R.~Rajkumar, ``{U}-connect: a low-latency
  energy-efficient asynchronous neighbor discovery protocol,'' in
  \emph{International Conference on Information Processingin Sensor Networks
  ({IPSN})}, 2010.

\bibitem{qiu:16}
Y.~Qiu, S.~Li, X.~Xu, and Z.~Li, ``Talk more listen less: Energy-efficient
  neighbor discovery in wireless sensor networks,'' in \emph{IEEE Conference on
  Computer Communications (INFOCOM)}, 2016.

\bibitem{meng:14}
T.~Meng, F.~Wu, and G.~Chen, ``On designing neighbor discovery protocols: A
  code-based approach,'' in \emph{IEEE Conference on Computer Communications
  (INFOCOM)}, 2014.

\bibitem{bleSpec50}
\mbox{Bluetooth SIG}, ``Specification of the {Bluetooth} system 5.0,'' December
  2016, volume 0, available via \mbox{\url{bluetooth.org}}.

\bibitem{AntSpec:14}
\mbox{Dynastream Innovations Inc.}, ``{ANT} message protocol and usage,'' 2014,
  revision 5.1, available via \mbox{\url{thisisant.com}}.

\bibitem{kindt:15}
P.~Kindt, M.~Saur, and S.~Chakraborty, ``Neighbor discovery latency in
  {BLE}-like protocols,'' \emph{{IEEE} Transactions on Mobile Computing},
  vol.~17, no.~3, pp. 617--631, 2018.

\bibitem{choi:11}
B.~J. {Choi} and X.~{Shen}, ``Adaptive asynchronous sleep scheduling protocols
  for delay tolerant networks,'' \emph{IEEE Transactions on Mobile Computing},
  vol.~10, no.~9, pp. 1283--1296, 2011.

\bibitem{jin:18}
X.~Meng, D.~Lin{-}Kit~Wong, B.~Leong, Z.~Wang, Y.~Don, and D.~Lu, ``Improving
  neighbor discovery by operating at the quantum scale,'' in \emph{{IEEE}
  International Conference on Mobile Ad Hoc and Sensor Systems {(MASS)}}, 2018.

\bibitem{kindt:17a}
P.~Kindt, D.~Yunge, G.~Reinerth, and S.~Chakraborty, ``Griassdi: Mutually
  assisted slotless neighbor discovery,'' in \emph{{ACM/IEEE} International
  Conference on Information Processing in Sensor Networks ({IPSN})}, 2017.

\bibitem{julien:17}
C.~Julien, C.~Liu, A.~L. Murphy, and G.~P. Picco, ``{BLEnd}: Practical
  continuous neighbor discovery for {Bluetooth Low Energy},'' in \emph{ACM/IEEE
  International Conference on Information Processing in Sensor Networks
  (IPSN)}, 2017.

\bibitem{loreti:19}
P.~Loreti and L.~Bracciale, ``Optimized neighbor discovery for opportunistic
  networks of energy constrained iot devices,'' \emph{IEEE Transactions on
  Mobile Computing}, vol.~19, no.~6, pp. 1387--1400, 2019.

\bibitem{li:14}
{Li}m D. and P.~{Sinha}, ``{RBTP}: Low-power mobile discovery protocol through
  recursive binary time partitioning,'' \emph{IEEE Transactions on Mobile
  Computing}, vol.~13, no.~2, pp. 263--273, 2014.

\bibitem{bracciale:16}
L.~{Bracciale}, P.~{Loreti}, and G.~{Bianchi}, ``The sleepy bird catches more
  worms: Revisiting energy efficient neighbor discovery,'' \emph{IEEE
  Transactions on Mobile Computing}, vol.~15, no.~7, pp. 1812--1825, 2016.

\bibitem{yang:15}
D.~{Yang}, J.~{Shin}, J.~{Kim}, and G.~{Kim}, ``Opeed: Optimal energy-efficient
  neighbor discovery scheme in opportunistic networks,'' \emph{Journal of
  Communications and Networks}, vol.~17, no.~1, pp. 34--39, 2015.

\bibitem{zheng:03}
R.~Zheng, J.~C. Hou, and L.~Sha, ``Asynchronous wakeup for ad hoc networks,''
  in \emph{ACM International Symposium on Mobile Ad Hoc Networking \& Computing
  (MobiHoc)}, 2003.

\bibitem{kindt:19}
P.~Kindt and S.~Chakraborty, ``On optimal neighbor discovery,'' in \emph{ACM
  Special Interest Group on Data Communication (SIGCOMM)}, 2019.

\bibitem{nrf51822:14}
\mbox{Nordic Semiconductor {ASA}}, ``{nRF51822} product spec. v3.1,'' 2014,
  available via \mbox{\url{nordicsemi.com}}.

\bibitem{liu:13}
J.~Liu, C.~Chen, Y.~Ma, and Y.~Xu, ``Adaptive device discovery in {Bluetooth
  Low Energy} networks,'' in \emph{IEEE Vehicular Technology Conference (VTC
  Spring)}, June 2013.

\bibitem{blessed:15}
\mbox{P. Borges et Al.}, ``{Bluetooth Low Energy} software stack for embedded
  devices ({BLESSED}),'' 2015,
  \mbox{\url{https://github.com/pauloborges/blessed}}.

\bibitem{kindt:13}
P.~H. Kindt, D.~Yunge, R.~Diemer, and S.~Chakraborty, ``Energy modeling for the
  {Bluetooth Low Energy} protocol,'' \emph{ACM Trans. Embed. Comput. Syst.},
  vol.~19, no.~2, Mar. 2020.

\bibitem{nordicSoftDev:19}
{Nordic Semiconductor ASA}, ``{S110 SoftDevice v8.0.0 API},'' 2015, available
  via
  \url{infocenter.nordicsemi.com/topic/com.nordic.infocenter.s110.api.v8.0.0/index.html}.

\end{thebibliography}
